\documentclass[fleqn,usenatbib]{mnras}

\usepackage{newtxtext,newtxmath}
\usepackage{amsmath,siunitx}	% Advanced maths commands
\usepackage{multicol}
\usepackage{multirow}
\usepackage{orcidlink} % package for including orchid IDs
\usepackage{pdflscape}	% Landscape pages
\usepackage{caption}
\usepackage{subcaption}
\usepackage{slashbox}
\usepackage[T1]{fontenc}

\DeclareRobustCommand{\VAN}[3]{#2}
\let\VANthebibliography\thebibliography
\def\thebibliography{\DeclareRobustCommand{\VAN}[3]{##3}\VANthebibliography}

\usepackage{graphicx}	% Including figure files
\usepackage{amsmath}	% Advanced maths commands
\title[Probabilistic classification of IR sources: YSOs]{Probabilistic Classification of Infrared-selected targets for SPHEREx mission: In search of YSOs}

\author[]{
K. Lakshmipathaiah\,\orcidlink{0000-0002-3889-0630}$^{1}$\thanks{Current affiliation: Space Astronomy Group, U. R. Rao Satellite Centre, Bangalore, 560037, India.},
S. Vig\,\orcidlink{0000-0002-3477-6021}$^{1}$\thanks{E-mail: sarita@iist.ac.in},
Matthew L. N. Ashby\,\orcidlink{0000-0002-3993-0745}$^{2}$\thanks{E-mail: mashby@cfa.harvard.edu}, 
Joseph L. Hora\,\orcidlink{0000-0002-5599-4650}$^{2}$,\newauthor 
Miju Kang\,\orcidlink{0000-0002-5016-050X}$^{3}$,
Rama Krishna Sai S. Gorthi\,\orcidlink{0000-0001-5021-0071}$^{4}$
\\
$^{1}$
Indian Institute of Space science and Technology (IIST), Thiruvananthapuram 695547, Kerala, India \\
$^{2}$Center for Astrophysics | Harvard \& Smithsonian, Optical and Infrared Astronomy Division, Cambridge, MA 01238, USA \\
$^{3}$Korea Astronomy and Space Science Institute, 776 Daedeokdaero, Yuseong, Daejeon 305-348, Republic of Korea\\
$^{4}$Indian Institute Of Technology–Tirupati, Tirupati 517506, India\\
}

\date{Accepted 2023 September 04. Received 2023 August 25; in original form 2022 November 20}

\pubyear{2023}

\begin{document}
\label{firstpage}
\pagerange{\pageref{firstpage}--\pageref{lastpage}}
\maketitle

\begin{abstract}
We apply machine learning algorithms to classify Infrared (IR)-selected targets for NASA's upcoming SPHEREx mission. In particular, we are interested in classifying Young Stellar Objects (YSOs), which are essential for understanding the star formation process.  Our approach differs from previous work, which has relied heavily on broadband color criteria to classify IR-bright objects, and are typically implemented in color-color and color-magnitude diagrams. However, these methods do not state the confidence associated with the classification and the results from these methods are quite ambiguous due to the overlap of different source types in these diagrams. Here, we utilize photometric colors and magnitudes from seven near and mid-infrared bands simultaneously and employ machine and deep learning algorithms to carry out probabilistic classification of YSOs, Asymptotic Giant Branch (AGB) stars, Active Galactic Nuclei (AGN) and main-sequence (MS) stars. Our approach also sub-classifies YSOs into Class I, II, III and flat spectrum YSOs, and AGB stars into carbon-rich and oxygen-rich AGB stars. We apply our methods to infrared-selected targets compiled in preparation for SPHEREx which are likely to include YSOs and other classes of objects. Our classification indicates that out of 8,308,384 sources, 1,966,340 have class prediction with probability exceeding 90\%, amongst which $\sim 1.7\%$ are YSOs, $\sim 58.2\%$ are AGB stars, $\sim 40\%$ are (reddened) MS stars, and $\sim 0.1\%$ are AGN whose red broadband colors mimic YSOs. We validate our classification using the spatial distributions of predicted YSOs towards the Cygnus-X star-forming complex, as well as AGB stars across the Galactic plane.
\end{abstract}


%%%%%%%%%%%%%%%%%%%%%%%%%%%%%%%%%%%%%%%%%%%%%%%%%%

%%%%%%%%%%%%%%%%% BODY OF PAPER %%%%%%%%%%%%%%%%%%

\section{Introduction}\label{sec:introduction}

The sources seen in the Infrared (IR) bands are diverse, and include objects such as the Young stellar objects (YSOs), reddened main-sequence (MS) stars, Asymptotic Giant Branch (AGB) stars, and Active Galactic Nuclei (AGN). Amongst these classes, YSOs represent the earliest phases of star formation embedded in their natal clouds, whilst the others represent mostly more evolved phases of stars (MS and AGB) or bright compact central regions of galaxies (AGN). 

Generally speaking, YSOs consist of a central accreting protostar or a pre-main sequence star surrounded by a circumstellar disk and/or envelope. The envelope and disk being cooler than the protostellar surface radiate in mid and far-infrared bands which is observed as an IR excess in comparison to MS stars \citep{Phil_Andre}. YSOs have been sub-classified based on the near and mid-infrared slope of their continuum emission, i.e., spectral index, with each sub-class corresponding to a certain stage in its evolution \citep{greene}. The infrared properties permit grouping of the YSOs into five broad categories, Class 0, Class I, flat~spectrum, Class II, and Class III YSOs, with Classes 0 and III respectively being the earliest and latest stages in the evolutionary sequence \citep{{LadaSpectralIndex},{Phil_Andre},{furlan}}. A statistical analysis of the observed YSO subclasses can, therefore provide significant clues regarding the time spent in each evolutionary stage \citep{{evans},{dunham}}. Identification of YSOs and their sub-classes from a large sample of IR-bright objects can be achieved by employing machine learning algorithms to segregate them from the contaminants that include (reddened) MS stars, AGB stars and AGN. Although our interest here is the investigation of YSOs, we emphasize that each set of classified objects, whether (reddened) MS stars, AGB stars or AGN can be analysed for their own respective characteristics.

%%%%%%%%%%%%%%%%%%%%%%%%%%%% FIG 1 Begins %%%%%%%%%%%%%%%%%%%%%%%%%%%%%%%%%%%%%%%%%%%

\begin{figure}
  \centering
  \includegraphics[width=0.45\textwidth]{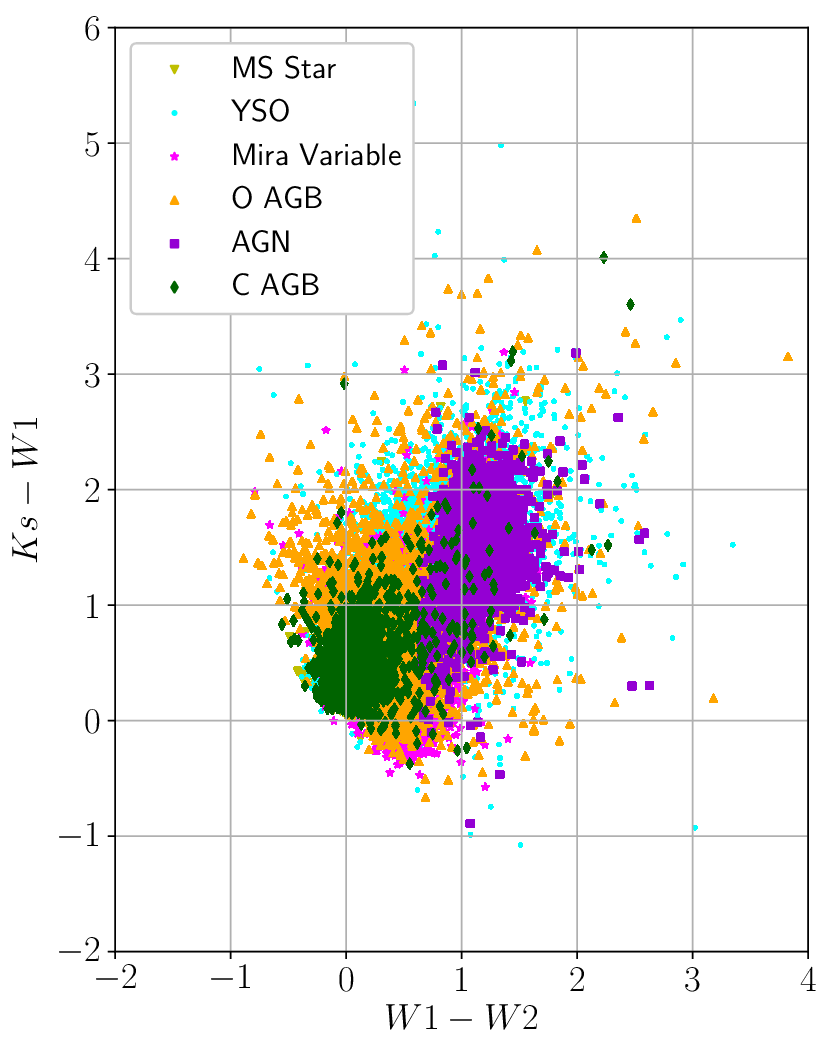}
  \caption{Overlapping distribution illustrated in $K_s-W1$ versus $W1-W2$ CCD. The distribution of different classes of objects overlaps in this and other alike CCDs or CMDs, making it difficult to distinguish among them. This requires the problem to be handled in higher dimensional space hoping to find separating hyper-plane in it.}
  \label{fig:Ks-W1_vs_W1-W2_CCD}
\end{figure}

%%%%%%%%%%%%%%%%%%%%%%%%%%%% FIG 1 Ends %%%%%%%%%%%%%%%%%%%%%%%%%%%%%%%%%%%%%%%%%%%

Large samples of YSOs have been conventionally identified using two or three photometric measurements, through different color-magnitude diagrams (CMD) and color-color diagrams (CCD) \citep[e.g.][]{gutermuth2009,fischer,grobschedl}. For example, \citet{koenig} use color-color diagrams constructed with \textit{Spitzer} Infrared Array Camera \citep[IRAC; ][]{Werner_2004,Fazio_2004} photometry to segregate YSOs from other object classes. One of the complications affecting these classification efforts is the tendency of contaminants to mimic YSO colors, leading to an overlap with the YSO distribution in the respective color-color and color-magnitude terrains \citep{{kuhn},{Lee_overlap_in_CCD_ref}}. The overlap of different non-YSO class objects with YSOs in $K_s-W1$ versus $W1-W2$ CCD is shown in Fig.~\ref{fig:Ks-W1_vs_W1-W2_CCD}. This traditional approach of classification essentially tries to build a rule-based system, wherein a set of conditions are imposed, and the objects satisfying them are classified into specific classes. One drawback of employing the CMDs and CCDs is that the classifications imposed by different authors are different, even if by small amounts. In addition, this is a deterministic classification scheme and does not admit the probablistic estimates.

Various researchers have attempted the classification of astronomical objects including YSOs using diverse machine learning algorithms. These methods utilise the supervised learning approach wherein labelled training data (with class labels for each source) are employed to train the machine learning models. We list a few such studies here. \citet{Vioque} used an Artificial Neural Network to train a classifier to probabilistically identify new pre-main sequence (PMS) candidates and classical Be stars in a sample of 4,150,983 sources resulting from cross-matching \textit{Gaia} DR2 \citep{{Gaia_mission_2016}, {Gaia_dr2_paper}}, Wide-field Infrared Survey Explorer \citep[AllWISE ; ][]{{allwise}, {allwise_catalog}}, IPHAS \citep{{Drew2005}, {Barentsen2014}} and VPHAS+ \citep{{Drew2014}}. They obtain 8470 new PMS candidates and 693 new classical Be candidates identified with probability exceeding 50\%, with completeness of $78.8\pm 1.4\%$ and $85.5\pm 1.2\%$, respectively. \citet{marton} tested several algorithms and found that the Random Forest (RF) classifier performed best. After training the RF classifier, \citet{marton} classified objects cross-matched between AllWISE and \textit{Gaia} DR2. More recently, \citet{kuhn} have used the Random Forest Classifier to identify 117,446 YSO candidates and 180,997 probable contaminants in a sample comprising \textit{Spitzer} sources with IR excess. The \citet{kuhn} catalog is one of the largest compilations of mid-IR-selected YSOs in the Galactic midplane. The spatial distribution and variability of the \citet{kuhn} candidate YSOs helps to validate their selection. \citet{chiu} have constructed the Spectrum Classifier of Astronomical Objects (SCAO) based on the Fully Connected Neural Network algorithm to classify any given source into regular stars, galaxies, or YSOs and used it to identify 129,219 candidate YSOs from \textit{Spitzer} Enhanced Imaging Products. The results show that source SEDs can be used to differentiate YSOs from other classes of objects, with the fluxes in the longer-wavelength IR bands being important features. \citet{cornu} extensively explored Artificial Neural Networks for YSO identification and classification. The authors conclude that the quality of the classifier mainly relies on the quality of training data used. Training sources having a wide range of feature values result in a diverse feature space, enabling the classifier trained upon it to more accurately classify the sources. \citet{miettinen} has explored multiple algorithms for the sub-classification of YSOs (Class 0, I and flat~spectrum). YSO sub-classification accuracy was found to mainly depend on the flux densities at \SI{3.6}{\micro\metre} and \SI{24}{\micro\metre}, and the accuracy can be improved using larger datasets and advanced ensemble methods. Unlike these studies, we have extended the exploration to deep learning algorithms (Convolutional Neural Networks) and employing an ensemble approach. We also attempt to incorporate uncertainty in input features during the model training and testing.

The Spectro-Photometer for the History of the Universe, Epoch of Reionization and Ices Explorer \citep[SPHEREx; ][]{SPHEREx_reference} is a NASA's upcoming mission with an anticipated launch no earlier than April 2025. Although this is an all-sky survey mission, for programmatic reasons 
spectra will be generated only for sources at mission-specified locations in the initial stages, i.e., at the coordinates given in a catalog supplied to the ground processing pipeline software. Hence, a target list is required for this purpose. As the name implies, one of the primary aims of the SPHEREx mission is to detect biogenic ices such as H$_2$O, CO$_2$, etc., along the lines-of-sight toward YSOs and protoplanetary disks, from a target list that has been assembled by \citet{Ashby_2023}. The SPHEREx target list includes IR-selected targets that are red, either due to intrinsic effects (YSOs / AGB stars) or due to extinction (e.g. MS stars).

Our objective in the present work is to classify the SPHEREx targets and identify candidate YSOs from other classes of objects. Like the earlier related work by \citet{blaum2019}, here we seek to classify SPHEREx targets so as to identify those which are likely to present absorption features due to ice species such as YSOs, and those which are essentially contaminants. Accordingly, we classify objects as YSOs, (reddened) MS stars, Asymptotic Giant Branch (AGB) stars, and Active Galactic Nuclei (AGN) using near and mid-infrared photometry in seven broad bands. 

Most of the machine learning based YSO catalogs in the literature, such as those of \citet{Vioque}, \citet{kuhn}, \citet{chiu} and \citet{cornu} are small compared to the SPHEREx target list. The all-sky catalog of \citet{marton} is relatively large and contains 103 million sources. However, the overlap between SPHEREx target list and \citet{marton} is only about 15\%. In addition, the input features of optical bands used by them would not be helpful for the classification of the earlier classes of YSOs. The near and mid-infrared bands are more suitable for this task as only the evolved YSOs or those with tenuous foreground medium are likely to be visible in optical. Thus, only a small fraction of the complete target list can be labelled by cross-matching with existing catalogs of machine learning YSO samples.

The machine learning based YSO classifiers in the literature use alternate and varied features for classification besides the Two Micron All-Sky Survey \citep[2MASS; ][]{2mass_catalog, 2mass} and AllWISE photometry. However, not all these features are available for all the SPHEREx targets $-$ \citet{Vioque} use photometry from 2MASS, AllWISE, \textit{Gaia} DR2, IPHAS and VPHAS+ catalogs, while \citet{marton} employ 2MASS, AllWISE, \textit{Gaia} DR2 photometry and Planck Dust Opacity. \citet{kuhn} utilise photometry from \textit{Spitzer}/IRAC, 2MASS, United Kingdom Infra-Red Telescope (UKIRT) Infrared Deep Sky Survey \citep[UKIDSS; ][]{Lawrence_2007} and the Visible and Infrared Survey Telescope for Astronomy (VISTA) Variables in the Vía Láctea survey \citep[VVV; ][]{Minniti_2010}, \citet{chiu} employ UKIDSS, \textit{Spitzer}/IRAC and MIPS bands photometry, and \citet{cornu} use photometry in \textit{Spitzer}/IRAC and MIPS bands. By construction, the SPHEREx targets are selected based on the 2MASS and AllWISE photometry that serves our purpose well as the IR bands are more suitable for classification of YSOs as compared to optical (IPHAS, VPHAS+, Gaia, etc).

We build upon the earlier efforts in three ways. (i) The existing machine learning based classifiers in the literature use any one of several types of machine learning algorithms. Unlike these, we employ multiple classification algorithms and combine these by making use of an ensemble classifier which provides a balanced classification with reasonable class-specific recovery rate. (ii) We employ a two-stage classification that yields more reliable results for the sub-classes of YSOs and AGB stars. (iii) We make use of photometric uncertainties in the classification process. Further, we explore convolutional neural networks for classification, which takes advantage of the correlation between photometry in adjacent bands for accurate classification. Our work also benefits from the recent publication of large catalogs that serve as training data. All these improvements are described in detail below. The adopted machine and deep learning methods take the assimilative information from seven near and mid-IR photometric bands simultaneously to learn the classification. The considered algorithms utilize the raw features (photometric colors and magnitudes) of a large set of labelled samples spanning different source types to learn and extract patterns of separability suitable for classifying seemingly non-separable classes in CCDs and CMDs.

To achieve our source classification goal, we (i) assembled a diverse training dataset, (ii) built a machine learning based probabilistic ensemble classifier using efficient and complimentary base classifiers, (iii) validated the classifier performance against known sources, (iv) probabilistically classified the sources in the SPHEREx target list, and (v) demonstrated the distribution of classified YSOs towards a molecular cloud complex. The organization of this paper is as follows. Section~\hyperref[sec:data_description]{2} of this paper describes the data used for training the machine learning models and the SPHEREx target list. Section~\hyperref[sec:methodology]{3} presents the different supervised machine learning algorithms explored and the proposed ensemble framework employed for the present analysis. In Section~\hyperref[sec:results]{4}, the results are analysed, which is followed by a discussion in Section~\hyperref[sec:exploration]{5}. The key findings are summarized in Section~\hyperref[sec:summary]{6}.

\section{Input Features and Data Description}
\label{sec:data_description}

\subsection{Input Features}

The main focus of this work is to classify sources in the SPHEREx target list of IR-bright sources. One of the primary goals of the SPHEREx mission is to identify and obtain spectra of ice-bearing YSOs and (reddened) MS stars, which will aid in understanding the origin of biogenic compounds at the early stages of planetary system formation. Thus, photometry in near and mid-IR bands would be helpful in classifying the targets and identifying candidate YSOs. We, therefore, employ photometric magnitudes and colors of sources as input features from the 2MASS and AllWISE catalogs for training and testing the machine learning models. The specific input features used for classification with the algorithms are described in detail in Sec.~\ref{sec:hyperparameter_tuning}.

\subsection{Data Description}

Here we follow an approach known as supervised machine learning, which relies on labelled data to train a machine learning algorithm, and then uses the trained algorithm to classify unknown objects in a collection of unlabelled data. The labelled data is split into training and validation sets, while the unlabelled data is referred to as the target list. The training set is used for building the machine learning model(s) and the validation set is used to evaluate the trained model and derive the confusion matrix and other performance metrics. The dataset to which the trained machine learning model(s) will be applied for its intended use, i.e., to classify the unlabelled samples, is called the target list. The training data is assembled from the literature as described below. The SPHEREx target list constitutes the unlabelled target list. 

\subsubsection{Training data}
\label{sec:training_data}
To train and validate the machine learning models, a set of labelled data has been assembled. Known as the training data, these were collated from the literature (including CDS and SIMBAD) describing known YSOs, and AGB stars from the Milky Way. The MS stars were selected to have types reported by SIMBAD in the range B0 - M8 with subtype spacings of 2, plus O6, O7, O8, and O9 stars, requiring that they have valid 2MASS and WISE $W1$ and $W2$ band detections. The MS stars in the color-selected target list are expected to be veiled by foreground dust and gas. We have, therefore applied an additional random extinction to the $\sim 17,000$ MS stars assembled, with an exponential distribution that closely matches the slope of the $A_V$ distribution of SPHEREx targets (not equal weighted but peaked near $A_V=2$\,mag and decreasing as $A_V$ increases), using the $A_{\lambda}/A_V$ ratios given by \citet{Wang_Chen_reddening_law}. Ten random extinctions were applied for every input star to construct a distribution of $\sim 170,000$ objects. Because of the 2MASS sensitivity, the maximum $A_V$ that has all bands 2MASS $J$, $H$, $K_s$ and AllWISE $W1$ and $W2$ bands is about $28-29$\,mag. The sample has about 155,000 objects with good photometry in all bands $J$ to $W2$. We then down-sampled this set to have $\sim 17,000$ (reddened) MS stars, in order to prevent a very high class imbalance. Amongst these training data MS stars, the stars of spectral class O, B, A, F, G, K, and M constitute 0.2\%, 8.8\%, 9.8\%, 25.0\%, 27.9\%, 18.1\% and 10.2\%, respectively. In addition to the Galactic infrared bright sources, probable extragalactic AGN contaminants were also retrieved. The references from where all these objects are taken are listed in Appendix \ref{sec:training_data_refs}. 

The 2MASS and AllWISE magnitudes were extracted from the published papers. In case the reference paper did not include the 2MASS and/or AllWISE entries of a source, the same were extracted from InfraRed Space Archive (IRSA) using the Gator search engine within a radius of \SI{1.5}{\arcsecond}. We found a few same-class duplicates amongst the training data collated from different catalogs, which were duly removed. In case the same source was found to be classified into broadly distinct classes by different authors, it was discarded from the training data due to ambiguity in the class label. Also, only those sources with photometric uncertainties less than 1\,mag in each band are considered for training. 

The source identification method on {\sl WISE} images is known to produce spurious photometric results, which is discussed in the literature \citep[e.g.][]{Koenig_Leisawitz_2014,Marton_2016,Silverberg_2018,marton}. Hence, following an approach similar to that in \citet{marton}, we intend to train two classifiers $-$ one using only photometry in 2MASS $J$, $H$, $K_s$, and AllWISE $W1$ and $W2$ bands (which we refer to henceforth as the $non-W3W4$ classifier) and another using photometry in 2MASS $J$, $H$, $K_s$, and AllWISE $W1$, $W2$, $W3$ and $W4$ bands (the $W3W4$ classifier). For training the $non-W3W4$ classifier, we consider only those sources with photometric quality flag \textit{ph\_qual =``A'' or ``B''} in 2MASS $J$ and $H$ bands, \textit{ph\_qual =``A''} in 2MASS $K_s$, AllWISE $W1$ and $W2$ bands. This results in a training data of size 53,638 sources of which 17,064 are YSOs, 17,064 are (reddened) MS stars, 4861 are AGN, 2207 are C AGB stars, 5567 are O AGB stars and 6875 are Mira Variables. For training the $W3W4$ classifier, in addition to the selection criteria applied to the $non-W3W4$ classifier training data we apply selection criteria to consider sources with real detections in $W3$ and $W4$ bands. For this, we require that \textit{ph\_qual =``A''}, the reduced chi-square \textit{$rchi2 < 10$} in AllWISE $W3$ and $W4$ bands, and $W4<7.5$\,mag. Further, we exclude targets satisfying $W1-W2<0.35$\,mag and $W3-W4>1.40$\,mag, since sources in this region were found to prominently have spurious photometry in $W3$ and $W4$ bands as determined from visual inspection. This gives us training data with a total of 25,133 sources, of which with 6135 are YSOs, 1289 are (reddened) MS stars, 4561 are AGN, 1606 are C AGB stars, 5167 are O AGB stars, and 6375 are Mira Variables. These are utilised for training the $W3W4$ classifier.

While building a machine learning model, it is crucial to have the training data samples similar to the target list samples. This ensures uniformity in defining the class boundaries in the model. To make the training data similar to the the SPHEREx target list, only sources brighter than 11.94\,mag in $W2$ and satisfying the color criterion: $H-W2>0.324$\,mag were considered. This is discussed in detail below.  

\subsubsection{Target List}
\label{sec:target_list}
% Each target in 
The SPHEREx target list (unlabelled data) includes near- and mid-IR photometry from 2MASS ($J$, $H$ and $K_s$), AllWISE ($W1, W2, W3$, and $W4$) and $Spitzer$/IRAC (3.6, 4.5, 5.8, and \SI{8.0}{\micro\metre}) wavebands. Not all sources have $Spitzer$/IRAC associations, hence we have used only 2MASS and AllWISE data. Several techniques were used to select sources expected to lie behind ever-greater levels of dust extinction (see \citet{Ashby_2023} for more details). The targets selected in the list are those that are bright enough for detection by SPHEREx with a high signal-to-noise ratio, by applying the magnitude criterion $W2<11.94$\,mag. Also, the color criterion: $H-W2>0.324$\,mag (or, if not detected in the $H$~band, $K_s-W2>0.55$\,mag) was applied to include reddened sources with $A_V>2$\,mag. 

For our classification, we have only considered targets that satisfy the selection criteria similar to that applied to training data, that is, targets with \textit{ph\_qual =``A'' or ``B''} in 2MASS $J$ and $H$ bands, \textit{ph\_qual =``A''} in 2MASS $K_s$, AllWISE $W1$ and $W2$ bands. Amongst them, sources with \textit{2MASS\_rd\_flg =0,9} were excluded as they correspond to either non-detection or a nominal detection. In addition, we found sources that were already present in the assembled training data. These sources were removed from the target list. Our final target list catalog contains 8,308,384 sources for classification. All these sources are classified using the $non-W3W4$ classifier.

Amongst the target list sources, 490,104 targets are found to satisfy criteria with valid detections in $W3$ and $W4$ bands (as described in Sec.~\ref{sec:training_data}) and these objects are also classified with the relevant training data under the $W3W4$ classification scheme. Overall, we adopt the classifications from the $W3W4$ classifier for this subset of targets having valid detection in $W3$ and $W4$ bands and those from the $non-W3W4$ classifier are considered for others. The justification of combining the classifications from two different classifiers is that the classifications when using the additional $W3$ and $W4$ band photometry are more favorable as more input features are available. Further, we compare the classifications from the two classifiers and find good agreement between the classifications where the two sets overlap (see Sec.~\ref{sec:comparison_of_classifications}).

\section{Classification Scheme} 
\label{sec:methodology}

This section describes the machine learning algorithms and methods adopted for classification. Instead of following the conventional approach of building a single classifier for simultaneous classification and sub-classification of objects, a two-stage classification scheme is used (the justification is given in Sec.~\ref{sec:2_stage_classification}). In the two-stage classification, the sources are first classified into broad classes and then similar classes of objects are sub-classified using auxiliary classifiers. In addition, we have used an ensemble classifier that relies on the output of the other classifiers. This ensemble classification, which along with two-stage classification technique demonstrably improves the reliability of the target classifications. We also incorporate uncertainties in input features during training and classification. 

\subsection{Two-Stage classification}
\label{sec:2_stage_classification}
Generally, in classification problems, a single algorithm is utilized to classify all the objects. In the present case, there are 8 output classes: Class I, flat~spectrum, Class II, Class III YSOs, (reddened) MS star, AGN, carbon-rich AGB (C~AGB) and oxygen-rich AGB (O~AGB) star. This includes both broadly different classes of objects (e.g. YSOs, AGN, AGBs) and sub-classes of the same class objects (e.g. different classes of YSOs). Trying to build a classifier to simultaneously classify and sub-classify the objects into one of the 8 output classes results in a significantly complex classifier. On the contrary, a two-stage classification scheme helps reduce this complexity by separating classification from sub-classification. The total number of parameters in a two-stage classifier can be higher than that in a one-stage classifier. Yet, merely increasing the number of parameters in a one-stage classifier does not help to achieve the desired simplification. Increasing the number of parameters in a one-stage classifier increases model complexity, which could lead to overfitting. Moreover, we observed an improvement in the mean cross-validation accuracy when we used a two-stage classification scheme in place of a one-stage classifier.

\begin{figure*}
   \centering
   \includegraphics[width=0.9\textwidth]{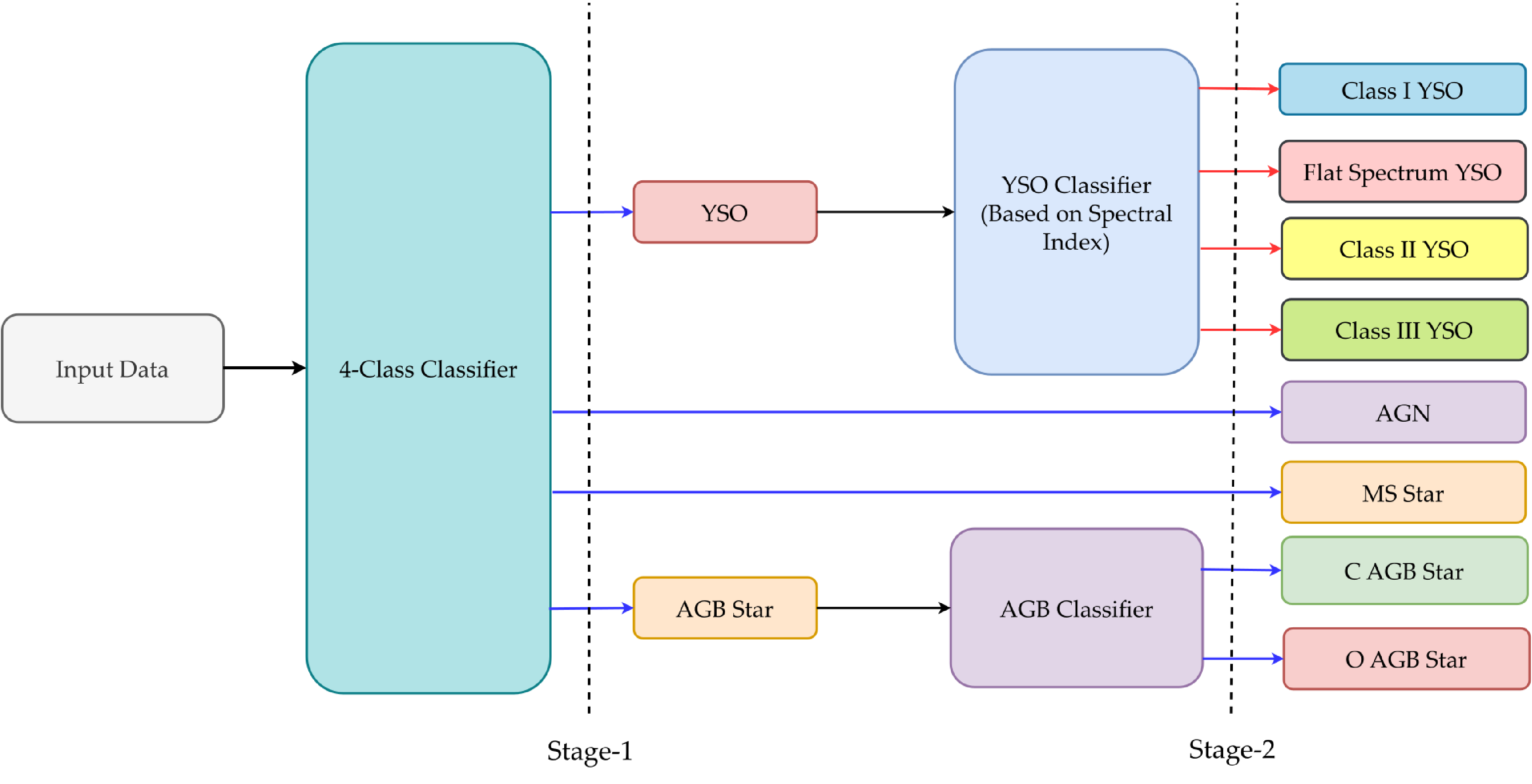}
   \caption{Visual depiction of the two-stage classifier. An ensemble of 100 RF, FNN and CNNs each, built for stage-1 classification as well as for AGB sub-classification. YSOs are sub-classified based on spectral index by adopting the uncertainty-based resampling method. Each spectral index is resampled 100 times and the fraction of times the sources is classified as each YSO sub-class is taken as the output probability. The classification based on spectral index is marked by red arrows and that based on machine learning algorithms is marked by blue arrows. The black arrows indicate the flow of input data in the classifier.}
   \label{fig:2_stage_classifier}
\end{figure*}

In our two-stage classification method, the first stage classifies sources into one of the broad classes, namely - YSO, (reddened) MS star, AGN, and AGB star. The second stage involves two classifiers, one for sub-classifying the YSOs (into Class~I, flat~spectrum, Class~II, Class~III) and the other for AGB stars (into C and O~AGB stars). The schematic of the two-stage classification is shown in Fig.~\ref{fig:2_stage_classifier}. In our two-stage classifier, the machine learning algorithms employing colors and magnitudes are used for the coarse classification (in stage-1) and AGB sub-classification (in stage-2), while spectral index \citep{greene} is used for YSO sub-classification. We estimate the spectral index by fitting a power law to the source SED considering the photometry in 2MASS $K_s$ and AllWISE $W1$, $W2$, $W3$ and $W4$ bands for targets with reliable photometry in $W3$ and $W4$ bands. For the rest, the spectral indices are determined using 2MASS $K_s$ and AllWISE $W1$, and $W2$ bands.

\subsection{Choice of classification algorithms}

We considered three supervised machine learning algorithms: (1) Random Forest Classifier \citep[RF; ][]{rf, decisionTree}, (2) Fully Connected Neural Network \citep[FNN; ][]{perceptron, backprop, bishop}, and (3) Convolutional Neural Network \citep[CNN; ][]{waibel, lenet, duda_hart_book, alexnet}. In addition, we constructed an ensemble classifier that utilizes these three base classifiers (RF, FNN, and CNN). Each source is assigned a class with the highest membership probability output by the ensemble classifier, since that represents the most likely class association for the source.

\subsection{Ensemble Classifier}

\begin{figure*}
   \centering
   \includegraphics[width=0.9\textwidth]{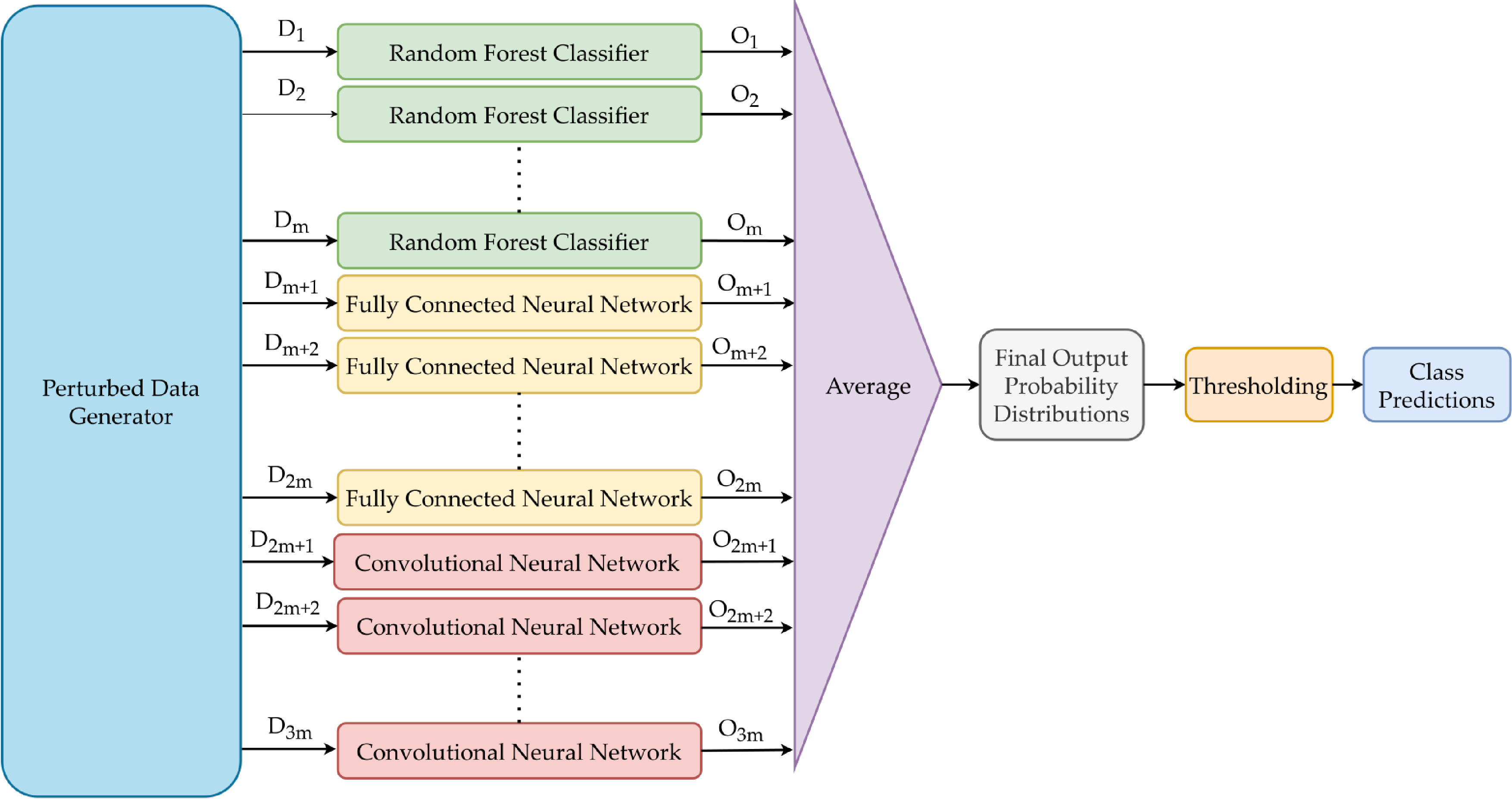}
   \caption{Visual depiction of the proposed ensemble classifier with the base classifiers RF, FNN and CNNs. The input data is perturbed using a Gaussian perturbation and passed through the different base classifiers for model training. $m$ (here, 100) classifiers of each type are trained and used to build the ensemble classifier bearing total of $3m$ classifiers whose output is averaged (here, giving equal weightage to each classifier output). The class with highest membership probability is considered as the output class if its associated probability exceeds the applied threshold.}
   \label{fig:1_stage_classifier}
\end{figure*}

All machine learning algorithms have weaknesses or limitations that are specific to them. To avoid their weaknesses yet benefit from their collective strengths, we constructed an ensemble classifier. The output of an ensemble classifier is based on the output probabilities of the individual base classifiers. The ensemble classifier using RF, FNN and CNN algorithms is depicted in Fig.~\ref{fig:1_stage_classifier}. A similar approach has been applied by \citet{bhavana} for the classification of brown dwarfs. More details describing our ensemble classifier and how it helps reduce the effects of overfitting are discussed in Appendix~\ref{sec:ensemble_over_fitting}.

\subsection{Uncertainty based resampling}
\label{sec:uncertainty_based_resampling}
One of the crucial assumptions while training the machine learning algorithms is that the data (input features and output class labels) are accurate, which is not always true given the uncertainty in the measurement of input features. An excellent method to incorporate uncertainties in inputs during classification is by resampling the data \citep{shy}. In this approach, each magnitude in the training data is resampled multiple times ($n=100$ in our case) from a Gaussian distribution centered on the measured magnitude, with a standard deviation equal to $\sqrt{2}$ times the associated photometric uncertainty. As described in Sec. 3 in \citet{shy}, the factor of $\sqrt{2}$ comes in due to the following considerations: (i) a uniform prior probability distribution function on the true value of magnitude, (ii) Gaussian distribution of observed magnitude centered on the (unknown) true magnitude, and (iii) having a standard deviation equal to the known photometric uncertainty in Bayesian analysis. Thus, $n$ perturbed training datasets are generated, and a classifier is fit to each of these perturbed sets resulting in $n$ trained classifiers. During classification, $n$ perturbed versions of the target list are generated following this approach. Each target list sample is passed through one of the trained classifiers for obtaining the class predictions. The $n$ outputs from the set of classifiers are then averaged to get the mean probability distribution over the output classes for each target. The target is labelled as belonging to the class with the highest output probability. 
This method helps to incorporate photometric uncertainties during classification effectively while providing the advantage of an ensemble approach. While considering the 2MASS photometry, we use the uncertainties listed in the \textit{msigcom} column of the 2MASS All-Sky Point Source Catalog, which accounts for the nightly zero point uncertainty. For AllWISE photometric uncertainties, we use the profile fit photometric uncertainty and the RMS instrumental zero point magnitudes for WISE All-Sky release \citep{AllWISE_zeropoint_err} added in quadrature.

%%%%%%%%%%%%%%%%%%%%%%%%%%%% FIG 2 Begins %%%%%%%%%%%%%%%%%%%%%%%%%%%%%%%%%%%%%%%%%%%

%%%%%%%%%%%%%%%%%%%%%%%%%%%% FIG 2 Ends %%%%%%%%%%%%%%%%%%%%%%%%%%%%%%%%%%%%%%%%%%%

%%%%%%%%%%%%%%%%%%%%%%%%%%%% FIG 3 Begins %%%%%%%%%%%%%%%%%%%%%%%%%%%%%%%%%%%%%%%%%%%

%%%%%%%%%%%%%%%%%%%%%%%%%%%% FIG 3 Ends %%%%%%%%%%%%%%%%%%%%%%%%%%%%%%%%%%%%%%%%%%%

\subsection{Hyperparameter tuning}
\label{sec:hyperparameter_tuning}

The assembled labelled data (training data) is split into two subsets called training and validation sets. The training set, as the name suggests, is used for model training. The performance of the trained classifier is then gauged on unseen samples included in the validation set. To mitigate selection effects, a $k$-fold cross-validation is performed. In this technique, the labelled data is shuffled and split into $k$ subsets. In each cross-validation, one of the subsets serves as validation set and the rest function together as a training set. This process of training and validation is repeated until each fold is used as validation set precisely once, leading to $k$ trained models. This process can be repeated multiple times, say $m$ by shuffling the data. This process results in a total of $m\times k$ trained models that can be used for testing. In our case, we performed 5-fold cross-validation 20 times to generate 100 models of each base classifier (in each stage of classification). We use uncertainty-based resampled versions of the labelled data for training and validation in each cross-validation.

%%%%%%%%%%%%%%%%%%%%%%%%%%%% Table 1 Begins %%%%%%%%%%%%%%%%%%%%%%%%%%%%%%%%%%%%%%%%%%%

\begin{table}
   \centering
   \caption{The $non-W3W4$ classifier stage-1 4-class classifier class-specific recall (recovery rate, in \%).}
   \begin{tabular}{|l||c|c|c|c|}
   \hline
   \backslashbox[43mm]{Classifier}{Class}
   &\makebox[2em]{AGB}&\makebox[2em]{AGN}&\makebox[2em]{MS}&\makebox[2em]{YSO}\\\hline\hline
   Random Forest Classifier & 97.5 & 94.5 & 95.8 & 89.9 \\
   \hline
   Fully Connected Neural Network & 97.2 & 96.9 & 96.0 & 87.2 \\
   \hline
   Convolutional Neural Network & 96.5 & 96.2 & 95.4 & 85.0 \\
   \hline
   Complete Ensemble Classifier & 97.4 & 96.3 & 96.1 & 88.0 \\
   \hline
\end{tabular}
   \label{tab:4_class_classifier_class_specific_recall}
\end{table}

%%%%%%%%%%%%%%%%%%%%%%%%%%%% Table 1 Ends %%%%%%%%%%%%%%%%%%%%%%%%%%%%%%%%%%%%%%%%%%%

%%%%%%%%%%%%%%%%%%%%%%%%%%%% Table 2 Begins %%%%%%%%%%%%%%%%%%%%%%%%%%%%%%%%%%%%%%%%%%%

\begin{table}
   \centering
   \caption{The $non-W3W4$ classifier stage-2 AGB classifier class-specific recall (recovery rate, in \%).}
   \begin{tabular}{|l||c|c|c|c|}
   \hline
   \backslashbox[43mm]{Classifier}{Class}
   &\makebox[3em]{C AGB}&\makebox[3em]{O AGB}\\\hline\hline
   Random Forest Classifier & 72.2 & 93.5 \\
   \hline
   Fully Connected Neural Network & 83.1 & 83.4 \\
   \hline
   Convolutional Neural Network & 76.7 & 84.6 \\
   \hline
   Complete Ensemble Classifier & 78.8 & 89.2 \\
   \hline
\end{tabular}
   \label{tab:AGB_class_classifier_class_specific_recall}
\end{table}

%%%%%%%%%%%%%%%%%%%%%%%%%%%% Table 2 Ends %%%%%%%%%%%%%%%%%%%%%%%%%%%%%%%%%%%%%%%%%%%

%%%%%%%%%%%%%%%%%%%%%%%%%%%% Table 3 Begins %%%%%%%%%%%%%%%%%%%%%%%%%%%%%%%%%%%%%%%%%%%

\begin{table}
   \centering
   \caption{The $W3W4$ classifier stage-1 4-class classifier class-specific recall (recovery rate, in \%).}
   \begin{tabular}{|l||c|c|c|c|}
   \hline
   \backslashbox[43mm]{Classifier}{Class}
   &\makebox[2em]{AGB}&\makebox[2em]{AGN}&\makebox[2em]{MS}&\makebox[2em]{YSO}\\\hline\hline
   Random Forest Classifier & 99.4 & 96.4 & 69.9 & 95.5 \\
   \hline
   Fully Connected Neural Network & 98.6 & 96.2 & 87.1 & 82.2 \\
   \hline
   Convolutional Neural Network & 93.3 & 95.1 & 77.5 & 74.1 \\
   \hline
   Complete Ensemble Classifier & 99.1 & 96.8 & 82.6 & 89.0 \\
   \hline
\end{tabular}
   \label{tab:4_class_classifier_class_specific_recall_W34}
\end{table}

%%%%%%%%%%%%%%%%%%%%%%%%%%%% Table 3 Ends %%%%%%%%%%%%%%%%%%%%%%%%%%%%%%%%%%%%%%%%%%%

%%%%%%%%%%%%%%%%%%%%%%%%%%%% Table 4 Begins %%%%%%%%%%%%%%%%%%%%%%%%%%%%%%%%%%%%%%%%%%%

\begin{table}
   \centering
   \caption{The $W3W4$ classifier stage-2 AGB classifier class-specific recall (recovery rate, in \%).}
   \begin{tabular}{|l||c|c|c|c|}
   \hline
   \backslashbox[43mm]{Classifier}{Class}
   &\makebox[3em]{C AGB}&\makebox[3em]{O AGB}\\\hline\hline
   Random Forest Classifier & 91.7 & 98.7 \\
   \hline
   Fully Connected Neural Network & 95.0 & 96.8 \\
   \hline
   Convolutional Neural Network & 93.1 & 96.1 \\
   \hline
   Complete Ensemble Classifier & 94.3 & 97.8 \\
   \hline
\end{tabular}
   \label{tab:AGB_class_classifier_class_specific_recall_W34}
\end{table}

%%%%%%%%%%%%%%%%%%%%%%%%%%%% Table 4 Ends %%%%%%%%%%%%%%%%%%%%%%%%%%%%%%%%%%%%%%%%%%%

%%%%%%%%%%%%%%%%%%%%%%%%%%%% Table 5 Begins %%%%%%%%%%%%%%%%%%%%%%%%%%%%%%%%%%%%%%%%%%%

\begin{table*}
   \centering
   \caption{The $non-W3W4$ classifier stage-1 4-class classifier 20 times repeated 5-fold cross-validation results.}
   \begin{tabular}{|p{4cm}|p{3cm}|p{1.75cm}|p{5.75cm}|}
   \hline
   \textbf{Classifier} & \textbf{Input Features} & \textbf{Accuracy} (\%) & \textbf{Details} \\
   \hline
   \hline
   Random Forest Classifier & All magnitudes and colors & $94.3$ & No. of decision trees: 100, min\_samples\_split: 5 \\
   \hline
   Fully Connected Neural Network & Colors and $W2$ & $93.6$ & Epochs: 500, Learning rate: 0.001 \\
   \hline
   Convolutional Neural Network & Colors and $W2$ & $92.5$ & Epochs: 500, Learning rate: 0.001 \\
   \hline
   Complete Ensemble Classifier & - & $93.9$ & - \\
   \hline
\end{tabular}
   \label{tab:4_class_classifier_validation_results}
\end{table*}

%%%%%%%%%%%%%%%%%%%%%%%%%%%% Table 5 Ends %%%%%%%%%%%%%%%%%%%%%%%%%%%%%%%%%%%%%%%%%%%

%%%%%%%%%%%%%%%%%%%%%%%%%%%% Table 6 Begins %%%%%%%%%%%%%%%%%%%%%%%%%%%%%%%%%%%%%%%%%%%

\begin{table*}
   \centering
   \caption{The $non-W3W4$ classifier stage-2 AGB classifier 20 times repeated 5-fold cross-validation results.}
   \begin{tabular}{|p{4cm}|p{3cm}|p{1.75cm}|p{5.75cm}|}
   \hline
   \textbf{Classifier} & \textbf{Input Features} & \textbf{Accuracy} (\%) & \textbf{Details} \\
   \hline
   \hline
   Random Forest Classifier & All magnitudes and colors & $87.4$ & No. of decision trees: 100, min\_samples\_split: 5 \\
   \hline
   Fully Connected Neural Network & Colors and $W2$ & $83.3$ & Epochs: 500, Learning rate: 0.002 \\
   \hline
   Convolutional Neural Network & Colors and $W2$ & $82.3$ & Epochs: 500,  Learning rate: 0.002  \\
   \hline
   Complete Ensemble Classifier & - & $86.2$ & - \\
   \hline
\end{tabular}
   \label{tab:AGB_class_classifier_validation_results}
\end{table*}

%%%%%%%%%%%%%%%%%%%%%%%%%%%% Table 6 Ends %%%%%%%%%%%%%%%%%%%%%%%%%%%%%%%%%%%%%%%%%%%

%%%%%%%%%%%%%%%%%%%%%%%%%%%% Table 7 Begins %%%%%%%%%%%%%%%%%%%%%%%%%%%%%%%%%%%%%%%%%%%

\begin{table*}
   \centering
   \caption{The $W3W4$ classifier stage-1 4-class classifier 20 times repeated 5-fold cross-validation results.}
   \begin{tabular}{|p{4cm}|p{3cm}|p{1.75cm}|p{5.75cm}|}
   \hline
   \textbf{Classifier} & \textbf{Input Features} & \textbf{Accuracy} (\%) & \textbf{Details} \\
   \hline
   \hline
   Random Forest Classifier & All magnitudes and colors & $96.4$ & No. of decision trees: 100, min\_samples\_split: 5 \\
   \hline
   Fully Connected Neural Network & Colors and $W2$ & $93.6$ & Epochs: 500, Learning rate: 0.001 \\
   \hline
   Convolutional Neural Network & Colors and $W2$ & $88.1$ & Epochs: 500, Learning rate: 0.001 \\
   \hline
   Complete Ensemble Classifier & - & $95.4$ & - \\
   \hline
\end{tabular}
   \label{tab:4_class_classifier_validation_results_W34}
\end{table*}

%%%%%%%%%%%%%%%%%%%%%%%%%%%% Table 7 Ends %%%%%%%%%%%%%%%%%%%%%%%%%%%%%%%%%%%%%%%%%%%

%%%%%%%%%%%%%%%%%%%%%%%%%%%% Table 8 Begins %%%%%%%%%%%%%%%%%%%%%%%%%%%%%%%%%%%%%%%%%%%

\begin{table*}
   \centering
   \caption{The $W3W4$ classifier stage-2 AGB classifier 20 times repeated 5-fold cross-validation results.}
   \begin{tabular}{|p{4cm}|p{3cm}|p{1.75cm}|p{5.75cm}|}
   \hline
   \textbf{Classifier} & \textbf{Input Features} & \textbf{Accuracy} (\%) & \textbf{Details} \\
   \hline
   \hline
   Random Forest Classifier & All magnitudes and colors & $97.0$ & No. of decision trees: 100, min\_samples\_split: 5 \\
   \hline
   Fully Connected Neural Network & Colors and $W2$ & $96.4$ & Epochs: 500, Learning rate: 0.002 \\
   \hline
   Convolutional Neural Network & Colors and $W2$ & $95.4$ & Epochs: 500,  Learning rate: 0.002  \\
   \hline
   Complete Ensemble Classifier & - & $96.9$ & - \\
   \hline
\end{tabular}
   \label{tab:AGB_class_classifier_validation_results_W34}
\end{table*}

%%%%%%%%%%%%%%%%%%%%%%%%%%%% Table 8 Ends %%%%%%%%%%%%%%%%%%%%%%%%%%%%%%%%%%%%%%%%%%%

One of the conventional approaches for training is to split the labelled data into 80:20 ratio, with the former fraction being used for training and the latter for validation. We employ a method of 5-fold cross-validation that is similar to the 80:20 data split for training, but with additional benefits that reduce bias during the split process itself. During the splitting of labelled data into training and validation sets, the class population in each set was proportional to that in the complete labelled data. To further reduce the bias, we shuffle the labelled data 20 times and do the 5-fold cross validation each time leading to $20 \times 5 = 100$ different splits of training and validation sets.

We train the RF, FNN, CNNs and the ensemble classifier using each of the 100 pairs of training and validation sets. The validation set output from each of the 100 RF, FNN and CNN are averaged and each source is associated with the class with highest membership probability. In addition to the individual comparison with the available class labels, these results are now also passed through the ensemble classifier. The output of the ensemble classifier is compared with the available class labels through the confusion matrix and other performance metrics. The normalized confusion matrix of the $non-W3W4$ ensemble classifiers in stages-1 and 2 are displayed in Tables~\ref{tab:4classes_ensemble} and \ref{tab:AGBclasses_ensemble}, and that of the $W3W4$ ensemble classifiers in stages-1 and 2 in Tables~\ref{tab:4classes_ensemble_W34} and \ref{tab:AGBclasses_ensemble_W34}, respectively.

Similar to conventional modeling, machine learning algorithms include various free parameters (called hyperparameters) that need to be optimized manually and are not learned by the model during training. A performance metric that is to be optimized is selected and the set of hyperparameters giving a high and stable metric score across cross-validations is considered as the optimal choice. Among the available metrics, we use accuracy score to optimize the hyperparameters since it gives an estimate of number of misclassified samples. The most common approaches of searching for optimal hyperparameters include \textit{grid search} \citep{grid_search} or \textit{random search} \citep{random_search}. In our case, we use both grid and random search to find the optimal hyperparameters for each machine learning algorithm. The methods employed to build different machine learning algorithms are explained below. The class-specific recall values for each base classifier and the complete ensemble classifier in stage-1 and 2 of the $non-W3W4$ classifiers are displayed in Tables~\ref{tab:4_class_classifier_class_specific_recall} and \ref{tab:AGB_class_classifier_class_specific_recall}, and that of the $W3W4$ classifiers are displayed in Tables~\ref{tab:4_class_classifier_class_specific_recall_W34} and \ref{tab:AGB_class_classifier_class_specific_recall_W34}, respectively. Unlike any of the three base classifiers the ensemble classifier provides a balanced classification with reasonable class-specific recovery rate ($\gtrsim 80\%$) across all the output classes. For instance, even though C~AGB recall of RF in the $non-W3W4$ AGB sub-classifier is 69.3\%, due to ensembling, the C~AGB recall for the complete $non-W3W4$ AGB sub-classifier is 77.8\%. Further, the choice of hyperparameters and the cross-validation accuracy of each base classifier in stage-1 and stage-2 (AGB sub-classifier) of the $non-W3W4$ and $W3W4$ classifiers are displayed in Tables~\ref{tab:4_class_classifier_validation_results} and \ref{tab:AGB_class_classifier_validation_results}, and Tables~\ref{tab:4_class_classifier_validation_results_W34} and \ref{tab:AGB_class_classifier_validation_results_W34}, respectively, along with the complete ensemble classifier's accuracy. The cross validation results are exhibited in Appendix~\ref{sec:cross-validation_results}.

The number of objects belonging to different classes in the assembled training data are quite different. Training with class imbalance data can lead to a model that has learned to classify the dominant class samples better than the small (dilute) class samples. To enable the model to learn to classify each class equally well, irrespective of their class population, we penalize the classifier more when a dilute class sample is misclassified. This was achieved by scaling the classification error of each sample by the inverse of its class frequency.

\subsubsection{Random Forest Classifier Hyperparameter Tuning}

The prominent hyperparameter of RF is the number of decision trees. The depth of the tree is regulated by specifying the threshold on the number of samples in any internal node for it to split. The results from the grid search showed that the RF performance is nearly the same when the number of decision trees exceeds 100. Hence, we opt for a random forest classifier with 100 decision trees whose internal nodes split if the number of samples in the node exceeds 5. We use the same hyperparameters for RF in both stages of the two-stage classifier.

Another important hyperparameter is the set of input features. We searched for the best combination of input features for RF training in the $non-W3W4$ classifier and found that using all adjacent colors ($J-H$, $H-K_s$, $K_s-W1$ and $W1-W2$) and magnitudes ($J$, $H$, $K_s$, $W1$ and $W2$) gives the best cross-validation accuracy. For RF training in the $W3W4$ classifier, best cross-validation accuracy is obtained using all adjacent colors ($J-H$, $H-K_s$, $K_s-W1$, $W1-W2$, $W2-W3$ and $W3-W4$) and magnitudes ($J$, $H$, $K_s$, $W1$, $W2$, $W3$ and $W4$), including photometry in $W3$ and $W4$ bands. This was also validated using ANOVA f-Test \citep{{oneWayAnova},{anova}}. The RF classifier was implemented using the {\tt sklearn} library in {\tt Python} language. To account for the class imbalance, the $class\_weight$ parameter in {\tt sklearn} module function for RF is set to `balanced', which internally modifies the weights to the inverse of class population. The normalized confusion matrices of RF in the $non-W3W4$ classifier's stages-1 and 2 and the $W3W4$ classifier's stages-1 and 2, along with the purity of classification for each class are displayed in Tables~\ref{tab:4classes_rf} and \ref{tab:AGBclasses_rf}, and Tables~\ref{tab:4classes_rf_W34} and \ref{tab:AGBclasses_rf_W34}, respectively.

\subsubsection{Neural Networks Hyperparameter Tuning: FNN and CNN}

%%%%%%%%%%%%%%%%%%%%%%%%%%%% FIG 4 Begins %%%%%%%%%%%%%%%%%%%%%%%%%%%%%%%%%%%%%%%%%%%

\begin{figure}
   \centering
   \includegraphics[width=0.45\textwidth]{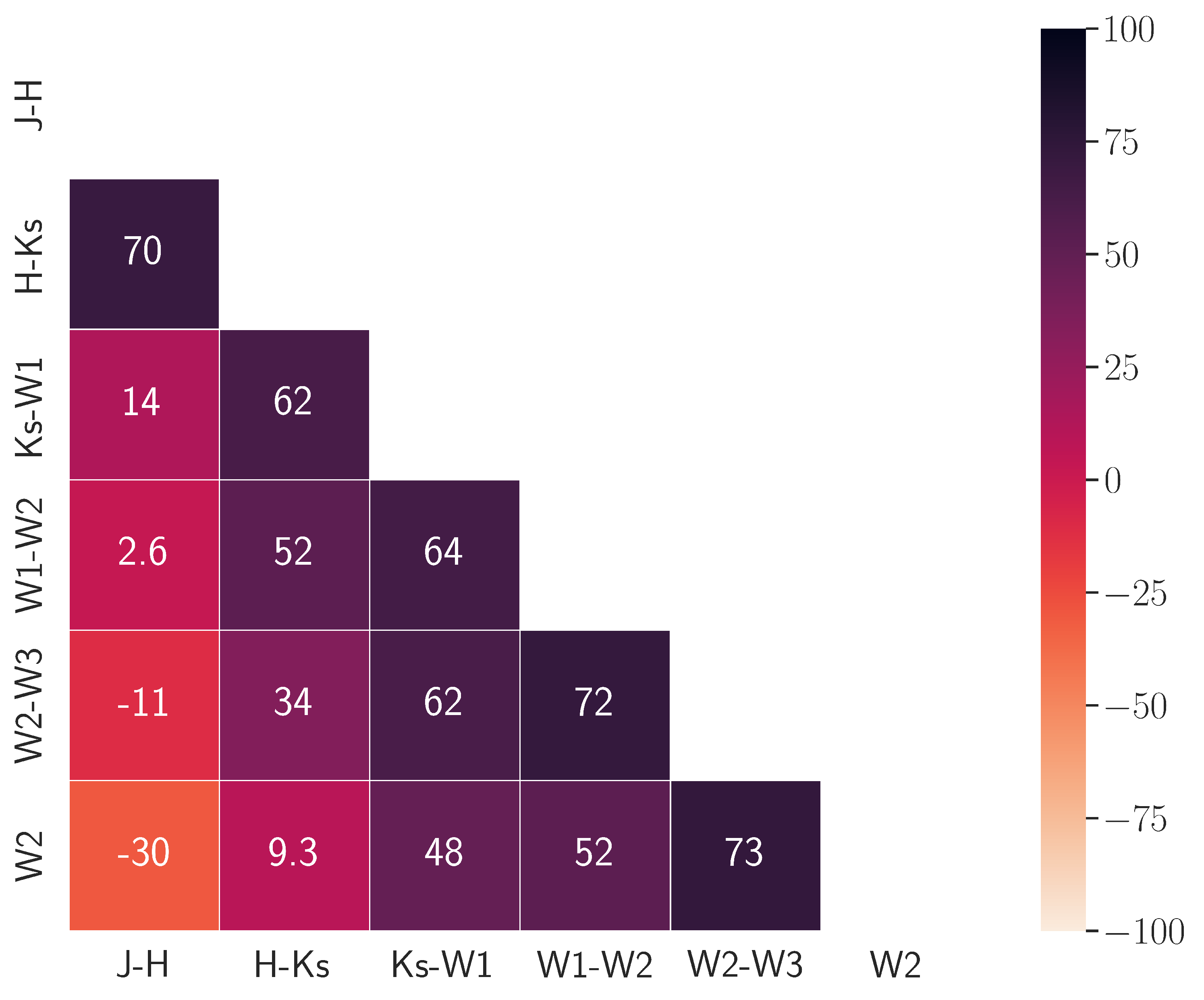}
   \caption{Pair-wise correlation matrix representing the Pearson's correlation coefficient for each pair of input features considered. The matrix cells are colorized according to their entry, which can range between -100 and +100\%. }
   \label{fig:corr_colors+W2}
\end{figure}

%%%%%%%%%%%%%%%%%%%%%%%%%%%% FIG 4 Ends %%%%%%%%%%%%%%%%%%%%%%%%%%%%%%%%%%%%%%%%%%%

The most important hyperparameters in neural networks are the input features, number of layers, number of neurons, learning rate, number of training epochs, batch size, optimizer, and the activation function. The architecture of the neural network is decided based on the results obtained from grid as well as random search. Various input features, such as the magnitudes, colors, and their combinations were tried for training each classifier, and the 5-fold cross-validation performance was noted. We found that the best results are obtained with $J-H$, $H-K_s$, $K_s-W1$ and $W1-W2$ colors and \textit{W2} magnitude as input features when using FNN and CNN in the $non-W3W4$ classifier. Whereas, for FNN and CNN in the $W3W4$ classifier, the best results are obtained with $J-H$, $H-K_s$, $K_s-W1$, $W1-W2$, $W2-W3$ and $W3-W4$ colors and \textit{W2} magnitude as input features. The correlation matrix, shown in Fig.~\ref{fig:corr_colors+W2}, reveals that the features are not tightly correlated. The correlation is $<75\%$ when we do not consider the adjacent features. There is a relatively higher correlation between adjacent features, which is preferable while using CNNs as the algorithm looks for local correlations. These factors support the choice of input features for FNN and CNN.

We used a 2-hidden layer FNN with 15 neurons in each hidden layer in the stage-1 (4-class) classifier (along with CNN and RF) in both the $non-W3W4$ and $W3W4$ classifiers. The second stage AGB sub-classifier includes a 1-hidden layer FNN with 20 hidden neurons in both the $non-W3W4$ and $W3W4$ classifiers. In the $non-W3W4$ ($W3W4$) classifier, a CNN with two (three) convolution layers was built using kernels of constant size (kernel size = 3) in each layer. The number of kernels in each subsequent layer was doubled, with the first layer having two and the last layer having four (eight) kernels. This results in a feature vector of length four (eight) at the end of the convolution layers (also called convolution block) which is followed by the auxiliary fully connected layers (also called fully connected block). The convolution layers are identical in both the stages-1 and 2 of the $non-W3W4$ and $W3W4$ classifiers. We tried kernels of size 2, 3 and 4 and found no significant difference in cross-validation performance. The fully connected block of CNN used in stage-1 of the $non-W3W4$ and $W3W4$ classifiers consists of 2~fully connected hidden layers with ten neurons in each hidden layer. On the other hand, the fully connected block of CNN in stage-2 AGB sub-classifier of the $non-W3W4$ and $W3W4$ classifiers consists of 1-hidden layer with ten neurons. We used the Rectified Linear Unit (ReLU) activation function \citep{relu} in all neural networks and trained each neural network (FNN and CNN) for 500 epochs. The FNN and CNN in the stage-1 classifier were trained with a learning rate of 0.001, whereas that in AGB sub-classifier with a learning rate of 0.002. The back-propagation \citep{backprop} was performed using the Adam optimizer \citep{Adam} with a mini-batch size of 4096. Softmax activation is used after the output layer for obtaining output probabilities. We included the batch-normalization layer \citep{batchnorm} at the end of each layer to standardize the inputs to the next layer. Cross-entropy loss function was used to estimate the classification error. We addressed the class imbalance by weighing each class's term in cross-entropy loss by the inverse of its class frequency. The normalized confusion matrices of FNN in the $non-W3W4$ classifier's stages-1 and 2 are displayed in Tables~\ref{tab:4classes_fcnn} and \ref{tab:AGBclasses_fcnn} and that in the $W3W4$ classifier's stages-1 and 2 in Tables~\ref{tab:4classes_fcnn_W34} and \ref{tab:AGBclasses_fcnn_W34}, respectively. The normalized confusion matrices of CNN in the $non-W3W4$ classifier's stages-1 and 2 are displayed in Tables~\ref{tab:4classes_cnn} and \ref{tab:AGBclasses_cnn} and that in the $W3W4$ classifier's stages-1 and 2 in Tables~\ref{tab:4classes_cnn_W34} and \ref{tab:AGBclasses_cnn_W34}, respectively. The FNN and CNN classifiers were implemented using the {\tt PyTorch} library in {\tt Python} language.

%%%%%%%%%%%%%%%%%%%%%%%%%%%% Table 9 Begins %%%%%%%%%%%%%%%%%%%%%%%%%%%%%%%%%%%%%%%%%%%

\begin{table*}
   \centering
   \caption{The numbers of classified sources as a function of probability threshold $P_{th}$ for the first stage of the two-stage classifier.}
\begin{tabular}{|c|c|c|c|c|c|c|c|c|c|}
\hline
\textbf{$P_{th}$} & \multicolumn{4}{c|}{\textbf{YSO}} & \multicolumn{2}{c|}{\textbf{AGB Star}}  & \textbf{AGN} & \textbf{MS} & \textbf{Total} \\
%\cline{4-5}
\cline{2-7}
& CI & FS & CII & CIII & C AGB & O AGB &  &  &  \\
\hline
\hline
\multirow{2}{*}{0.00} & \multicolumn{4}{c|}{616604} & \multicolumn{2}{c|}{2442394} & \multirow{2}{*}{6912} & \multirow{2}{*}{5242474} & \multirow{2}{*}{8308384} \\
\cline{2-7}
& 5231 & 8816 & 132946 & 469611 & 2247487 & 194907 & & &  \\ \hline 

\multirow{2}{*}{0.50} & \multicolumn{4}{c|}{368359} & \multicolumn{2}{c|}{2203369} & \multirow{2}{*}{6647} & \multirow{2}{*}{4893915} & \multirow{2}{*}{7472290} \\
\cline{2-7}
& 4993 & 8485 & 105877 & 249004 & 2011942 & 191427 & & & \\ \hline 

\multirow{2}{*}{0.75} & \multicolumn{4}{c|}{90060} & \multicolumn{2}{c|}{1677068} & \multirow{2}{*}{3753} & \multirow{2}{*}{3512392} & \multirow{2}{*}{5283273} \\
\cline{2-7}
& 3062 & 5778 & 55169 & 26051 & 1504630 & 172438 & & & \\ \hline 

\multirow{2}{*}{0.80} & \multicolumn{4}{c|}{70645} & \multicolumn{2}{c|}{1525387} & \multirow{2}{*}{3248} & \multirow{2}{*}{2922346} & \multirow{2}{*}{4521626} \\
\cline{2-7}
& 2557 & 4937 & 46914 & 16237 & 1359489 & 165898 & & & \\ \hline 

\multirow{2}{*}{0.85} & \multicolumn{4}{c|}{52329} & \multicolumn{2}{c|}{1373569} & \multirow{2}{*}{2656} & \multirow{2}{*}{2027137} & \multirow{2}{*}{3455691} \\
\cline{2-7}
& 2035 & 3981 & 37759 & 8554 & 1216235 & 157334 & & & \\ \hline 

\multirow{2}{*}{0.90} & \multicolumn{4}{c|}{34062} & \multicolumn{2}{c|}{1144165} & \multirow{2}{*}{2005} & \multirow{2}{*}{786108} & \multirow{2}{*}{1966340} \\
\cline{2-7}
& 1366 & 2855 & 26534 & 3307 & 1000354 & 143811 & & & \\ \hline 

\end{tabular}
   \label{tab:1_stage_threshold_stats}
\end{table*}

%%%%%%%%%%%%%%%%%%%%%%%%%%%% Table 9 Ends %%%%%%%%%%%%%%%%%%%%%%%%%%%%%%%%%%%%%%%%%%%

\section{Results}
\label{sec:results}

%%%%%%%%%%%%%%%%%%%%%%%%%%%% FIG 5 Begins %%%%%%%%%%%%%%%%%%%%%%%%%%%%%%%%%%%%%%%%%%%

\begin{figure}
   \centering
   \includegraphics[width=0.45\textwidth]{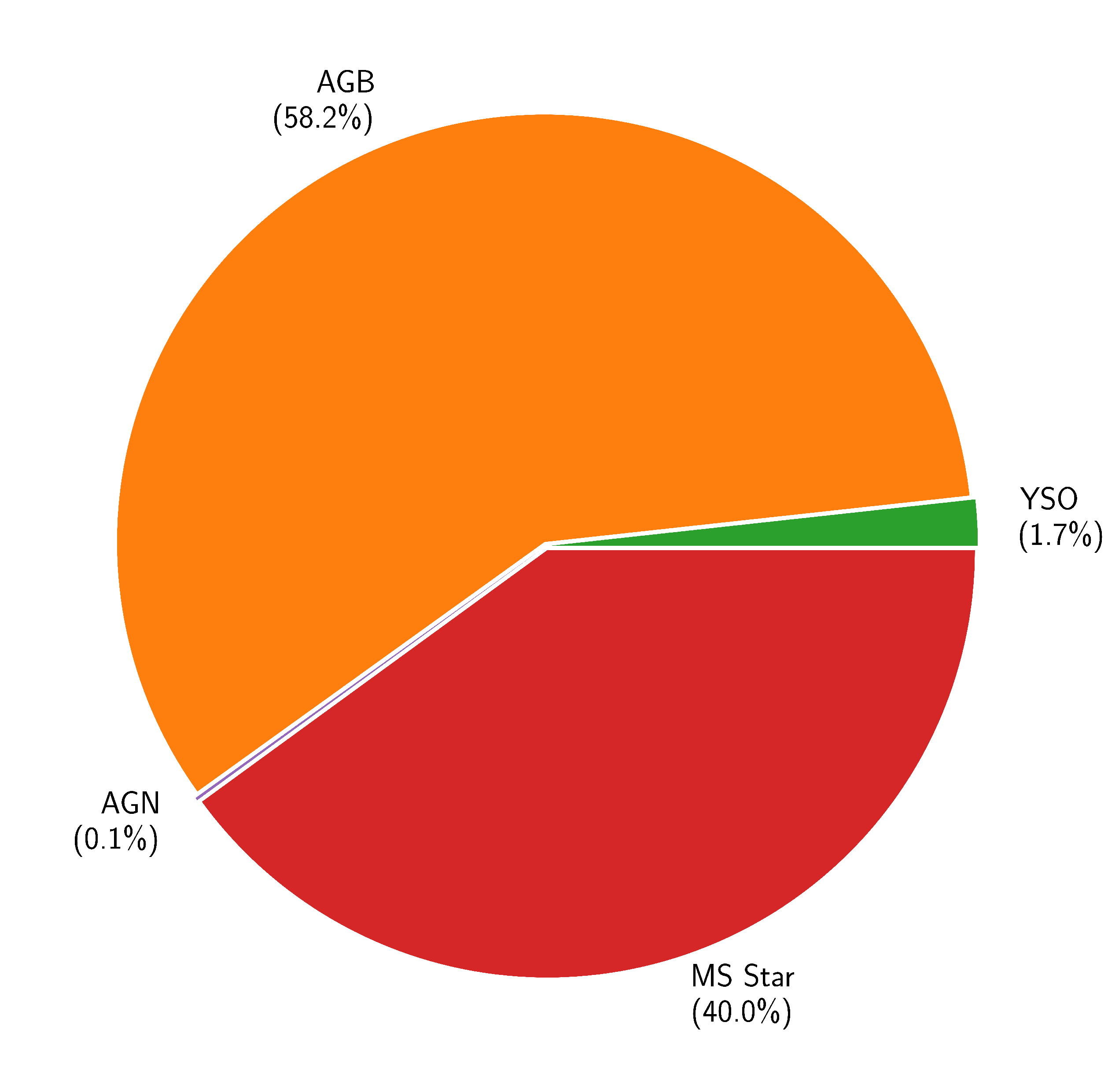}
   \caption{Pie chart of class population of target list obtained based on stage-1 classifier's predictions. The percentage in brackets indicate the class population percentage in the target list. The class population is given in the row corresponding to 90\% probability threshold in Table~\ref{tab:1_stage_threshold_stats}.}
   \label{fig:4classes_ensemble_test_pie_chart_with_percentages}
\end{figure}

%%%%%%%%%%%%%%%%%%%%%%%%%%%% FIG 5 Ends %%%%%%%%%%%%%%%%%%%%%%%%%%%%%%%%%%%%%%%%%%%

%%%%%%%%%%%%%%%%%%%%%%%%%%%% FIG 6 Begins %%%%%%%%%%%%%%%%%%%%%%%%%%%%%%%%%%%%%%%%%%%

\begin{figure}
   \centering
   \includegraphics[width=0.45\textwidth]{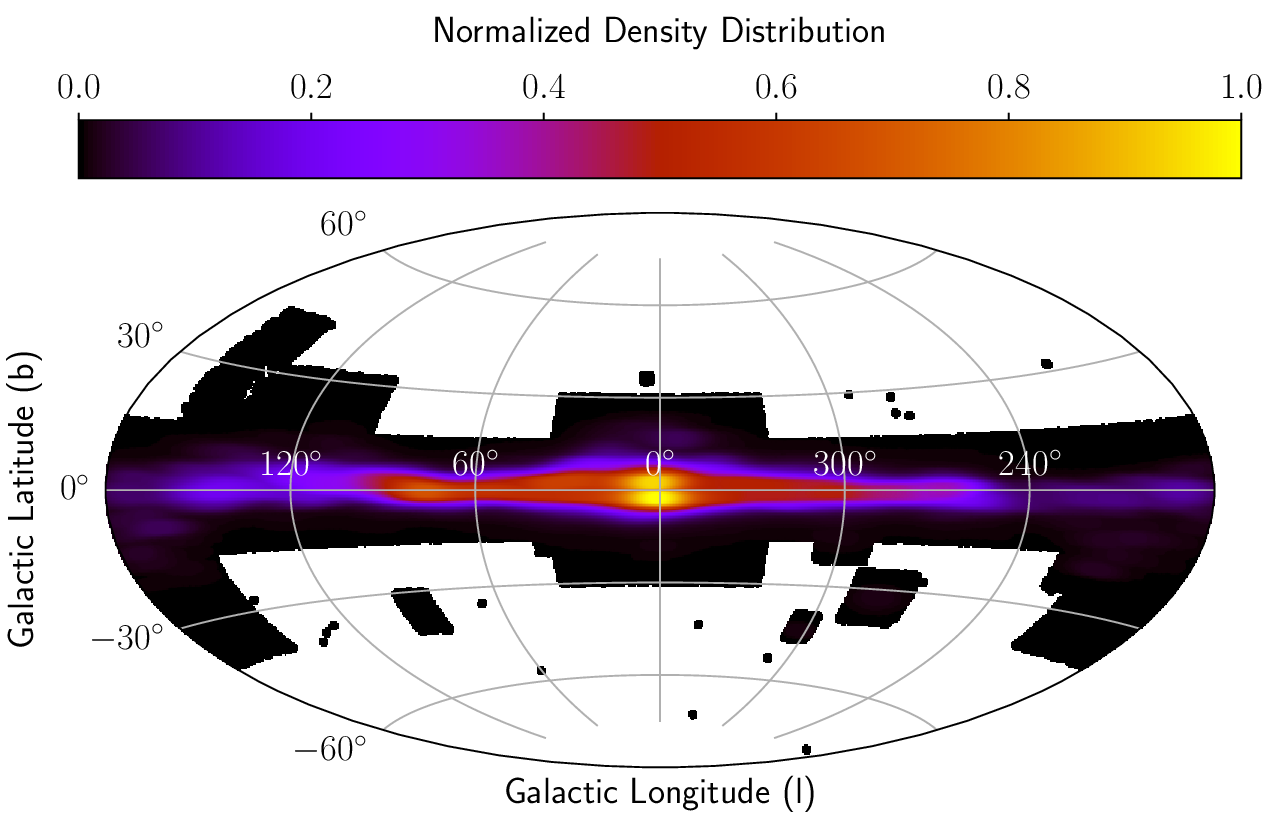}
   \caption{Plot of spatial distribution of 1,966,340 SPHEREx targets classified with probability exceeding 90\% in first stage of two-stage classifier on the Mollweide projection of the sky plane. The regions of sky bearing SPHEREx targets are colored using the coloring scheme displayed above the plot and those regions not bearing the SPHEREx targets are left blank.}
   \label{fig:pth_0.9_all_targets}
\end{figure}

%%%%%%%%%%%%%%%%%%%%%%%%%%%% FIG 6 Ends %%%%%%%%%%%%%%%%%%%%%%%%%%%%%%%%%%%%%%%%%%%

%%%%%%%%%%%%%%%%%%%%%%%%%%%% FIG 7 Begins %%%%%%%%%%%%%%%%%%%%%%%%%%%%%%%%%%%%%%%%%%%

\begin{figure}
   \centering
   \includegraphics[width=0.45\textwidth]{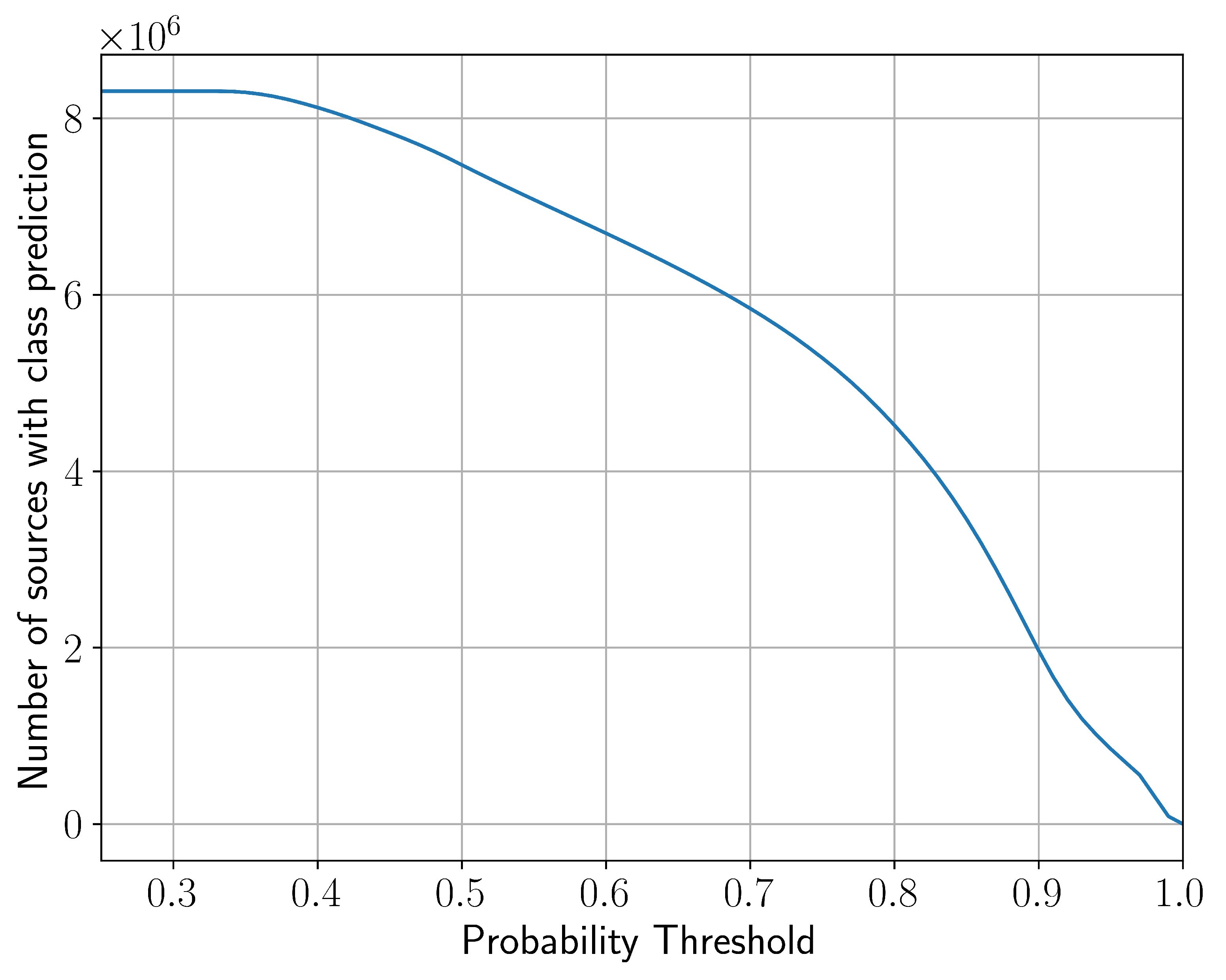}
 \caption{Plot of variation in the number of classified targets as a function of probability threshold applied on the output of stage-1 classifier. The lower limit of the x-axis is set to 0.25, since that is the worst possible output probability ($\frac{1}{4}\times 100\% = 25\%$), which constitutes a case where the classifier does not favor any classification higher than the others.}

 \label{fig:4classes_num_srcs_vs_pth}
\end{figure}

%%%%%%%%%%%%%%%%%%%%%%%%%%%% FIG 7 Ends %%%%%%%%%%%%%%%%%%%%%%%%%%%%%%%%%%%%%%%%%%%

%%%%%%%%%%%%%%%%%%%%%%%%%%%% FIG 8 Begins %%%%%%%%%%%%%%%%%%%%%%%%%%%%%%%%%%%%%%%%%%%

\begin{figure*}
   \centering
   \includegraphics[width=\textwidth]{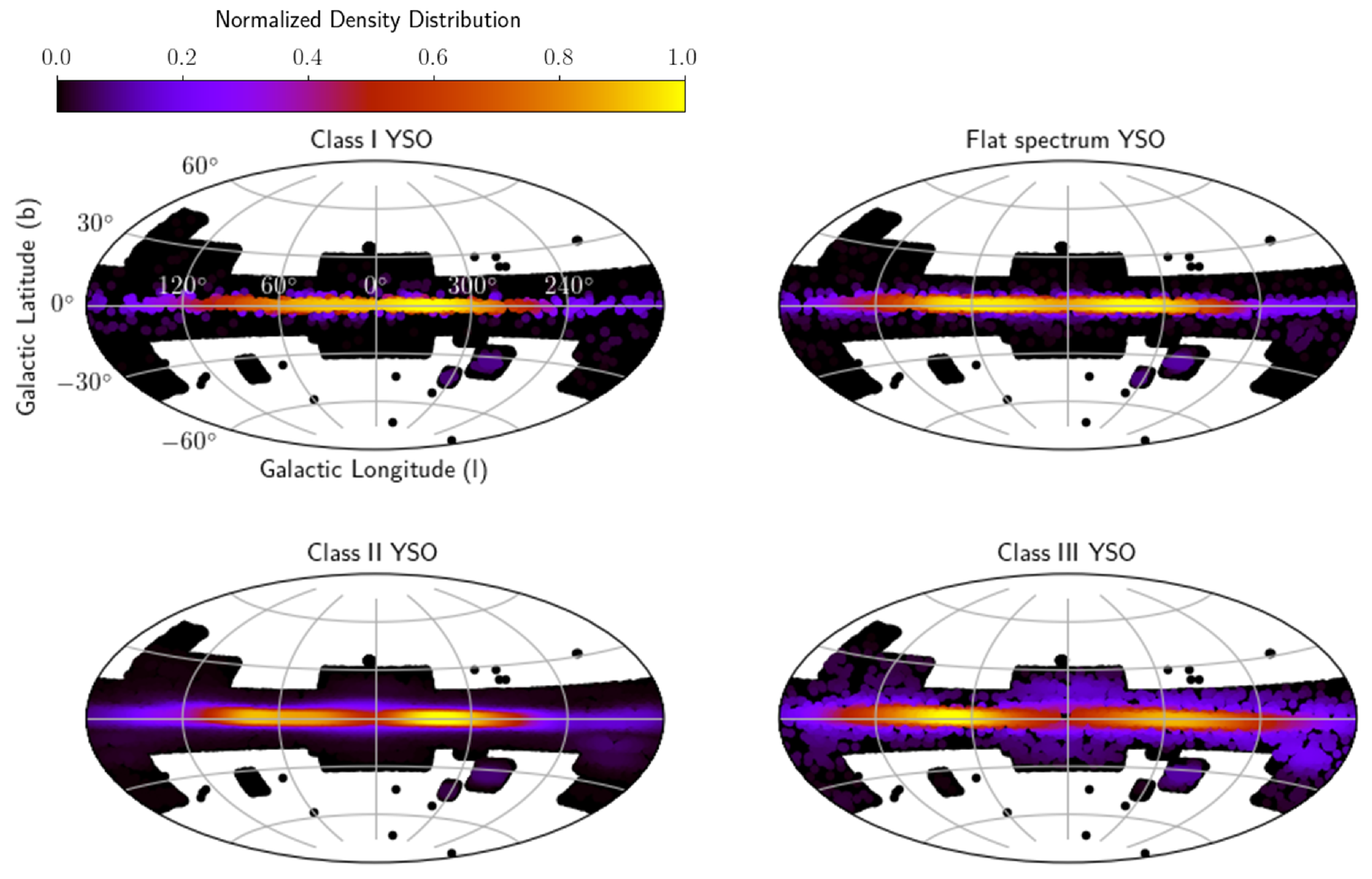}
   \caption{Plot of spatial distribution of different classes of YSOs in the target list on the Mollweide projection of the sky plane. There are 1366 Class~I, 2855 flat~spectrum, 26534 Class~II and 3307 Class~III YSOs identified with probability exceeding 90\% in first stage of two-stage classifier. These objects are identified mainly close to the Galactic mid-plane and some higher latitude molecular clouds. Regions occupied by SPHEREx targets are colorized using the coloring scheme displayed at the top of the plot and those not occupied by SPHEREx targets are left blank.}
   \label{fig:pth_0.9_combined_YSOs}
\end{figure*}

%%%%%%%%%%%%%%%%%%%%%%%%%%%% FIG 8 Ends %%%%%%%%%%%%%%%%%%%%%%%%%%%%%%%%%%%%%%%%%%%

The class prediction for each source in the target list is obtained using the two-stage ensemble classifier. We use the class with the highest probability to classify the source. The pie-chart in Fig.~\ref{fig:4classes_ensemble_test_pie_chart_with_percentages} shows the class distribution of the target list classified with probability exceeding 90\%, with the class labels as predicted by the stage-1 (4-class) ensemble classifier. From the target list, we find that 1,966,340 sources have class labels predicted with an output membership probability exceeding 90\%. Their spatial distribution is displayed in Fig.~\ref{fig:pth_0.9_all_targets}. We observe that the targets are mainly concentrated close to the Galactic plane and are also distributed in higher latitude molecular clouds and Magellanic clouds. The class population at this probability threshold can be viewed in Table~\ref{tab:1_stage_threshold_stats}. Of these sources, we find that 1.7\% are predicted as YSOs, 58.2\% as AGB stars, 40.0\% as (reddened) MS stars, and 0.1\% as AGN. Of the YSOs, we find that 4.0\% are Class I, 8.4\% are flat~spectrum, 77.9\% are Class II, and 9.7\% are Class III YSOs. Amongst the AGB stars, we find that 87.4\% are C~AGB while the rest 12.6\% as O~AGB stars.

Fig.~\ref{fig:4classes_num_srcs_vs_pth} shows the total number of classified targets as a function of probability threshold for the two-stage classification scheme. The number of sources under consideration drops with an increase in threshold probability. We see a steep decline in the number of classified sources when the probability threshold exceeds $\approx40\%$. The probability associated with classification depends on the distance from the class separation boundaries. This implies that a large majority of sources are lying near the separation boundaries of classes in the feature space or the regions overlap, leading to a low value of output probabilities. The lowest value of expected class probability is 25\%, corresponding to equal probabilities of all classes in stage-1. In our case, we find that the lowest output classification probability is 32.2\%.

%%%%%%%%%%%%%%%%%%%%%%%%%%%% FIG 9 Begins %%%%%%%%%%%%%%%%%%%%%%%%%%%%%%%%%%%%%%%%%%%

\begin{figure*}
   \centering
   \includegraphics[width=\textwidth]{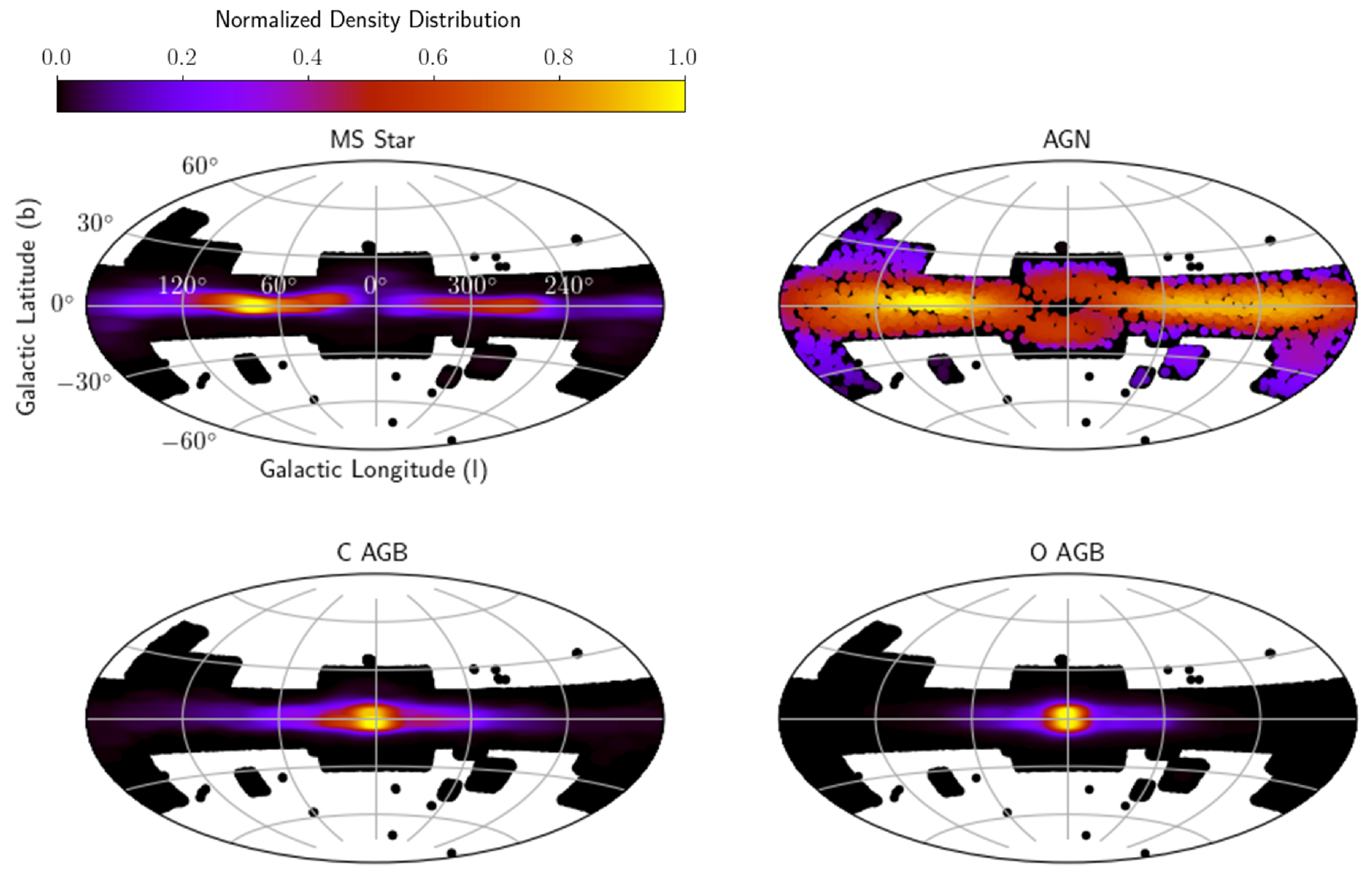}
   \caption{Plot of spatial distribution of (reddened) MS stars, AGN and AGB stars in the target list on the Mollweide projection of the sky plane. There are 786,108 (reddened) MS stars, 2005 AGN, 1,000,354 C~AGB and 143,811 O~AGB stars identified with probability exceeding 90\% in first stage of two-stage classifier. Regions occupied by SPHEREx targets are colorized using the coloring scheme displayed at the top of the plot and those not occupied by SPHEREx targets are left blank.}
   \label{fig:pth_0.9_combined_nonYSOs}
\end{figure*}

%%%%%%%%%%%%%%%%%%%%%%%%%%%% FIG 9 Ends %%%%%%%%%%%%%%%%%%%%%%%%%%%%%%%%%%%%%%%%%%%

Table~\ref{tab:1_stage_threshold_stats} gives two rows for each probability threshold (applied on stage-1 classifier output probability), one with the number of objects obtained from the stage-1 classifier, and the second row with the number of objects from the sub-classifier for each category within the class. If requried, the membership probability at each stage can be considered while applying the threshold. For example, a class label (after sub-classification) can be considered as reliable if $P(Class|Inputs)>P_{th}$ and $P(Sub-class|Class,Inputs)>P_{th}$. In other words, the classifier output at each stage is expected to exceed the threshold membership probability for being called a reliable prediction. Employing a threshold of 90\% on the output of both stages results in classification of 686,101 sources.

We examine the spatial distribution of the YSOs from the target list across the Galaxy. Fig.~\ref{fig:pth_0.9_combined_YSOs} shows the YSOs distributions after applying a threshold of 90\% on the membership probability in stage-1 of classification. The YSOs are found to densely populate the Galactic mid-plane with YSOs in different evolutionary stages having more or less similar Galactic distribution. We find that the Class II YSOs have the high density close to Galactic mid-plane. YSOs are also prominently identified towards the higher latitude molecular clouds such as Large Magellanic Cloud (LMC) and Small Magellanic Cloud (SMC), indicating active star~formation in these regions. The lines-of-sight towards these high latitude clouds are also found to have a significant population of (reddened) MS stars. Moreover, since the YSOs are sub-classified based on spectral index, this implies that the choice of photometric bands for estimating spectral index does affect the YSO sub-classification, and in-turn the relative population of YSO subclasses. Based on our YSO sub-classification, we find 1366 Class I, 2855 flat~spectrum, 26,534 Class II YSOs and 3307 Class III YSOs with probability exceeding 90\%. Here, we find that the population of Class II YSOs is nearly an order of magnitude higher than that of other YSO subclasses. Overall, most of the YSOs are more densely populated in the inner Galaxy ($|l|\lesssim90^\circ$) than in the outer Galaxy. This is in line with expectations since most of the young stars are likely to be found within the Milky Way Galaxy. The region within Galactic longitude of $\pm90^\circ$ mainly includes Milky Way sources, those within the Galactic radius of 8\,kpc (inner Galaxy). Whereas, the region outside this Galactic longitude range covers sources in the outer Galaxy or sources not associated with the Milky Way Galaxy. As it is more likely to find young stellar objects in dense regions of interstellar medium in the Milky Way, we find a high population density of YSOs in the inner Galaxy, which is clearly exhibited in the Galactic distribution plots of YSO subclasses.

Fig.~\ref{fig:pth_0.9_combined_nonYSOs} indicates that the (reddened) MS stars are found to populate the $60^\circ$ wide strip symmetric about the Galactic mid-plane, with their population density decreasing with an increase in the Galactic latitude from $0^\circ$ to $\pm90^\circ$. Owing to the minimum extinction criterion of the SPHEREx target list, the identified (reddened) MS stars are likely to be located behind the interstellar medium of the Milky Way Galaxy (which causes extinction) and hence, are identified prominently. Also, not many (reddened) MS stars are located outside Galactic longitude of $\pm 90^\circ$, since the degree of interstellar extinction is lower in the outer Galaxy relative to that in the inner Galaxy. The high density of (reddened) MS stars is found to be within Galactic longitude of $\pm 90^\circ$. The region within Galactic longitude of $\pm 90^\circ$ mainly includes Milky Way sources, whereas the AGN are identified away from the centre of the Milky Way Galaxy. This is in-line with expectations because extragalactic objects are less likely to be detected along the Galactic mid-plane owing to high extinction by dense clouds.  The C and O~AGB stars spatial distribution reveals that both the source types are prominently located in the inner Galaxy ($|l|\lesssim90^\circ$) and towards the Galactic center. In the target list, 1,000,354 C~AGB and 143,811 O~AGB stars are identified with probability exceeding 90\%, which means the population of C~AGB stars is nearly 7 times that of O~AGB stars in the target list. These sources are also found to be present towards higher latitude clouds such as LMC. In terms of absolute numbers, since the population of C~AGB stars is greater than that of O~AGB stars, we are likely to find more C~AGB stars than O~AGB stars in any given region included in the SPHEREx target list. 

The predicted class labels of the SPHEREx targets along with stages-1 and 2 classification probabilities and the classifier used (the $non-W3W4$ or $W3W4$) for classifying each source is included in the SPHEREx target list of ice sources (SPLICES) catalog, which is made available in electronic format through IRSA.

%%%%%%%%%%%%%%%%%%%%%%%%%%%% FIG 10 Begins %%%%%%%%%%%%%%%%%%%%%%%%%%%%%%%%%%%%%%%%%%%

\begin{figure*}
   \centering
   \includegraphics[width=\textwidth]{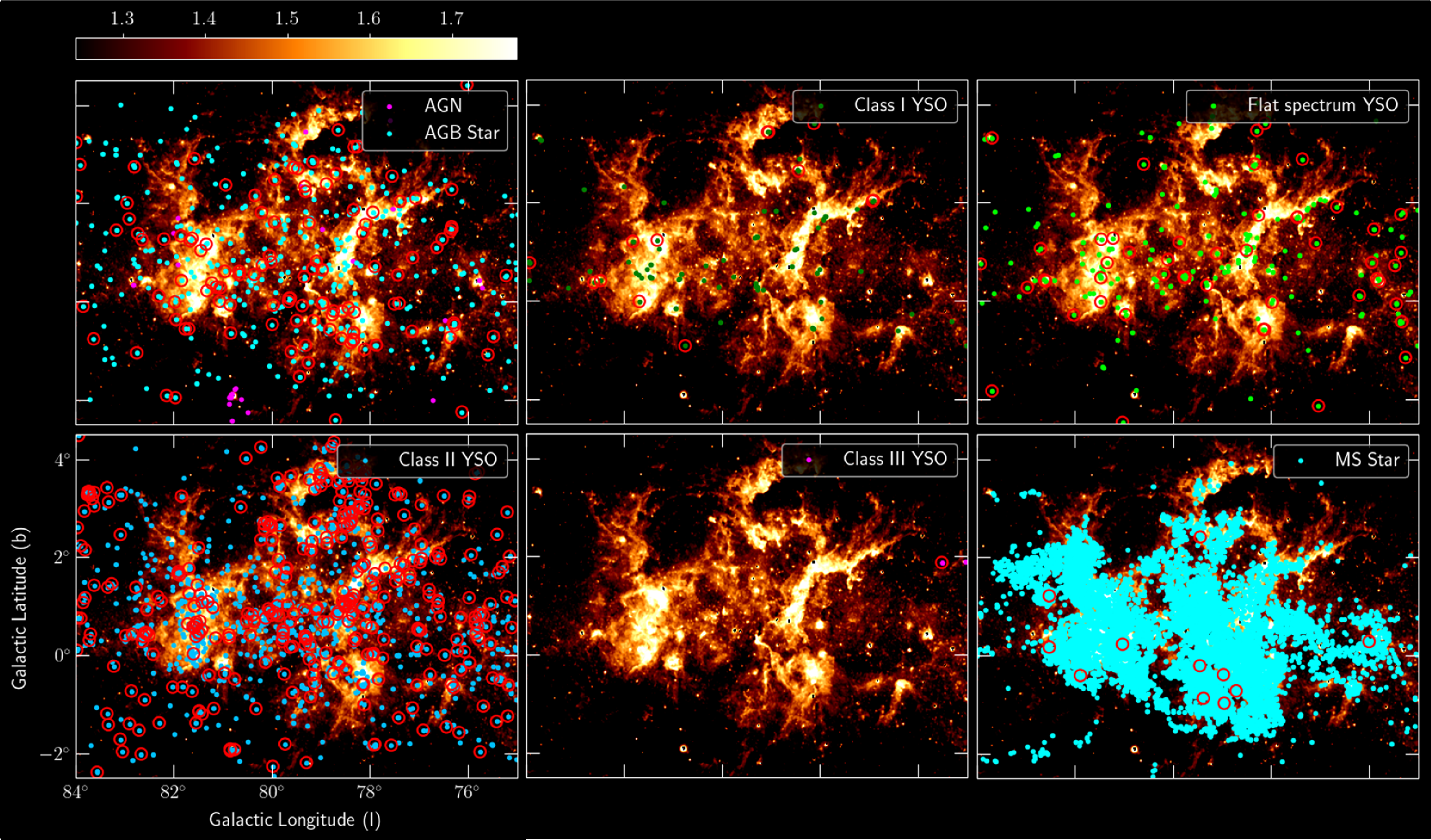}
   \caption{Spatial distribution of targets in the Cygnus-X region classified with probability exceeding 90\%; likely target type indicated by in-panel labels. The background image in every panel is the median-filtered {\sl WISE} {\sl W1} mosaic with color bar indicating relative surface brightness. The horizontal and vertical axes indicate Galactic longitude and latitude, respectively. The targets filtered based on the parallax criteria (those plotted in the box plot of distances in previous slide), are highlighted in red circles in the above figure. The targets filtered based on the parallax criterion are highlighted in red circles.}
\label{fig:cygnus_x_region}
\end{figure*}

%%%%%%%%%%%%%%%%%%%%%%%%%%%% FIG 10 Ends %%%%%%%%%%%%%%%%%%%%%%%%%%%%%%%%%%%%%%%%%%%

\subsection{Comparison of classifications from the $non-W3W4$ and $W3W4$ ensemble classifiers}
\label{sec:comparison_of_classifications}

Here we describe our comparison of the class predictions made by our two ensemble classifiers, the $W3W4$ and $non-W3W4$ classifier. For this analysis, only SPHEREx targets classified by both ensemble classifiers are considered. Among them, only those sources for which the class is predicted with a membership probability exceeding 90\% by both the classifiers are considered, which results in 373,542 targets. Comparing the class labels for these sources from the two ensemble classifiers, we see quite a good agreement between the two classifications.

%%%%%%%%%%%%%%%%%%%%%%%%%%%% Table 10 Begins %%%%%%%%%%%%%%%%%%%%%%%%%%%%%%%%%%%%%%%%%%%

\begin{table}
    \centering
    	\caption{Comparison matrix of classifications from the $non-W3W4$ and $W3W4$ classifiers.}
    \label{tab:classification_comparison_matrix}
\begin{tabular}{l|l|c|c|c|c|}
\multicolumn{2}{c}{}&\multicolumn{4}{c}{$non-W3W4$} \\
\cline{2-6}
& Class & AGB & AGN & MS star & YSO \\
\cline{2-6}
\multirow{4}{*}{\rotatebox[origin=c]{90}{$W3W4$}} & AGB & 370,391 & 0 & 0 & 0 \\ %\multicolumn{1}{c|}{}
\cline{2-6}
& AGN & 0 & 1409 & 0 & 0 \\
\cline{2-6}
& MS star & 0 & 0 & 1 & 0 \\
\cline{2-6}
& YSO & 0 & 0 & 0 & 1741 \\
\cline{2-6}
\end{tabular}
\end{table}

%%%%%%%%%%%%%%%%%%%%%%%%%%%% Table 10 Ends %%%%%%%%%%%%%%%%%%%%%%%%%%%%%%%%%%%%%%%%%%%

As can be seen from the comparison matrix in Table~\ref{tab:classification_comparison_matrix}, the classifications match for all considered sources. Thus, we find that with the same membership probability threshold, the class predictions by our two ensemble classifiers are in-line with each other. This implies that the classifications with limited photometry (that is, when not using $W3$ and $W4$ bands) are in-line with that obtained when all bands of 2MASS and AllWISE with good photometry in each band are considered, indicating reliability of our final classifications.

\section{Spatial distribution of sources}
\label{sec:exploration}

In the present section, we validate our results using spatial distributions of YSOs and AGB stars.

\subsection{YSOs towards Cygnus-X}

%%%%%%%%%%%%%%%%%%%%%%%%%%%% FIG 11 Begins %%%%%%%%%%%%%%%%%%%%%%%%%%%%%%%%%%%%%%%%%%%

\begin{figure}
   \centering
   \includegraphics[width=0.45\textwidth]{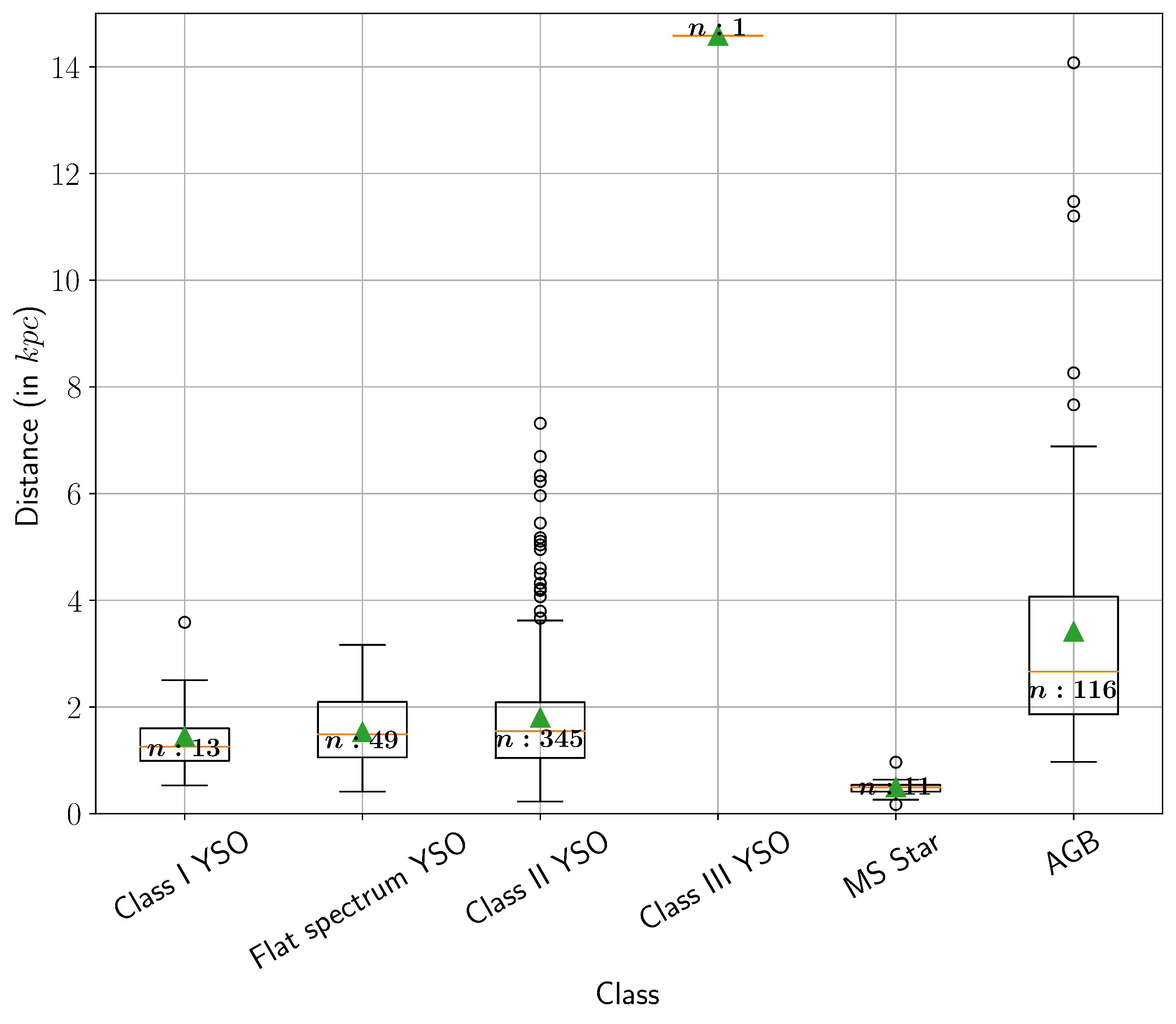}
   \caption{Box plot of distances to different classes of objects in Cygnus-X region. The mean distance to each class of objects is marked with a green triangle and the median distance using an orange horizontal line within the corresponding box.}
   \label{fig:distance_box_plot_0.9}
\end{figure}

%%%%%%%%%%%%%%%%%%%%%%%%%%%% FIG 11 Ends %%%%%%%%%%%%%%%%%%%%%%%%%%%%%%%%%%%%%%%%%%%

We extract YSOs towards a star-forming complex Cygnus-X from the SPHEREx target list and analyse their spatial distributions. The distance to Cygnus-X is estimated to be $1.40\pm0.08$\,kpc using parallax and proper motions of water masers towards this complex \citep{rygl}. In order to include only those sources that are likely to be members of the Cygnus-X cloud, we select sources that are highly reddened with visual extinction $A_V>$12\,mag, by applying the color criterion $H-W2>1.5$\,mag. Then, we apply a probability threshold of 90\% to the stage-1 identifications and get 10,508 sources, of which 0.8\% are Class I, 1.9\% are flat~spectrum, and 11.0\% are Class II YSOs, 82.2\% are (reddened) MS stars, 3.8\% are AGB stars, and 0.2\% are AGN. Merely two sources are identified as Class III YSOs. Fig.~\ref{fig:cygnus_x_region} shows the locations of all classified sources in Cygnus-X overlaid on the AllWISE {\sl W1} mosaic of the field. The spatial distributions of YSOs closely follows the filamentary structures within the cloud, while the (reddened) MS stars spatial spread indicates the distribution of dense molecular material in Cygnus-X region, since these sources are selected to have $A_V>12$\,mag. By contrast, the AGB stars and AGN are distributed relatively evenly throughout the field.  Unlike the YSOs their distributions show no evident relationship to the filamentary structure so prominently revealed in {\sl WISE} {\sl W1}. 

We have considered the \textit{Gaia}-DR3 \citep{{Gaia_mission_2016}, {Gaia_dr3_catalog}, {Gaia_dr3_2022}} counterparts to the sources in the Cygnus-X region with threshold probability exceeding 90\%, from the target list. We find 408 YSOs with \textit{Gaia} associations having relative uncertainty in the parallax value less than 50\%. We find that most of the YSOs except one Class III YSO are located at a similar distance as the cloud, which can be visualised from the box plot shown in Fig.~\ref{fig:distance_box_plot_0.9}. The AGB stars, on the other hand, are situated farther away from the cloud and constitute background objects. The average distance of the Class I, flat~Spectrum and Class II YSOs towards the cloud is $1.76\pm0.56$\,kpc, which matches the distance determined by the proper motions quite well. Thus, we infer that the Cygnus-X is an active star~forming region with 1440 of the sources towards it classified as YSOs with a probability exceeding 90\%. 

The coincidence of the YSO distributions with the filaments in Cygnus-X, the agreement of the distances, and the lack of evidence for correlation between AGN and AGB stars, and filaments, all support the accuracy of the attributions by the two-stage classifier.

\subsection{C and O AGB star distribution about the Galactic plane}

We validate our AGB sub-classifier predictions by analysing the C and O~AGB stars in the target list. For this, we consider a total of 621,632 C~AGB and 41,732 O~AGB stars from the catalog, that are predicted with a probability exceeding 90\% by both stage-1 and stage-2 classifiers.

%%%%%%%%%%%%%%%%%%%%%%%%%%%% FIG 12 Begins %%%%%%%%%%%%%%%%%%%%%%%%%%%%%%%%%%%%%%%%%%%

\begin{figure}
    \centering
    \includegraphics[width=0.45\textwidth]{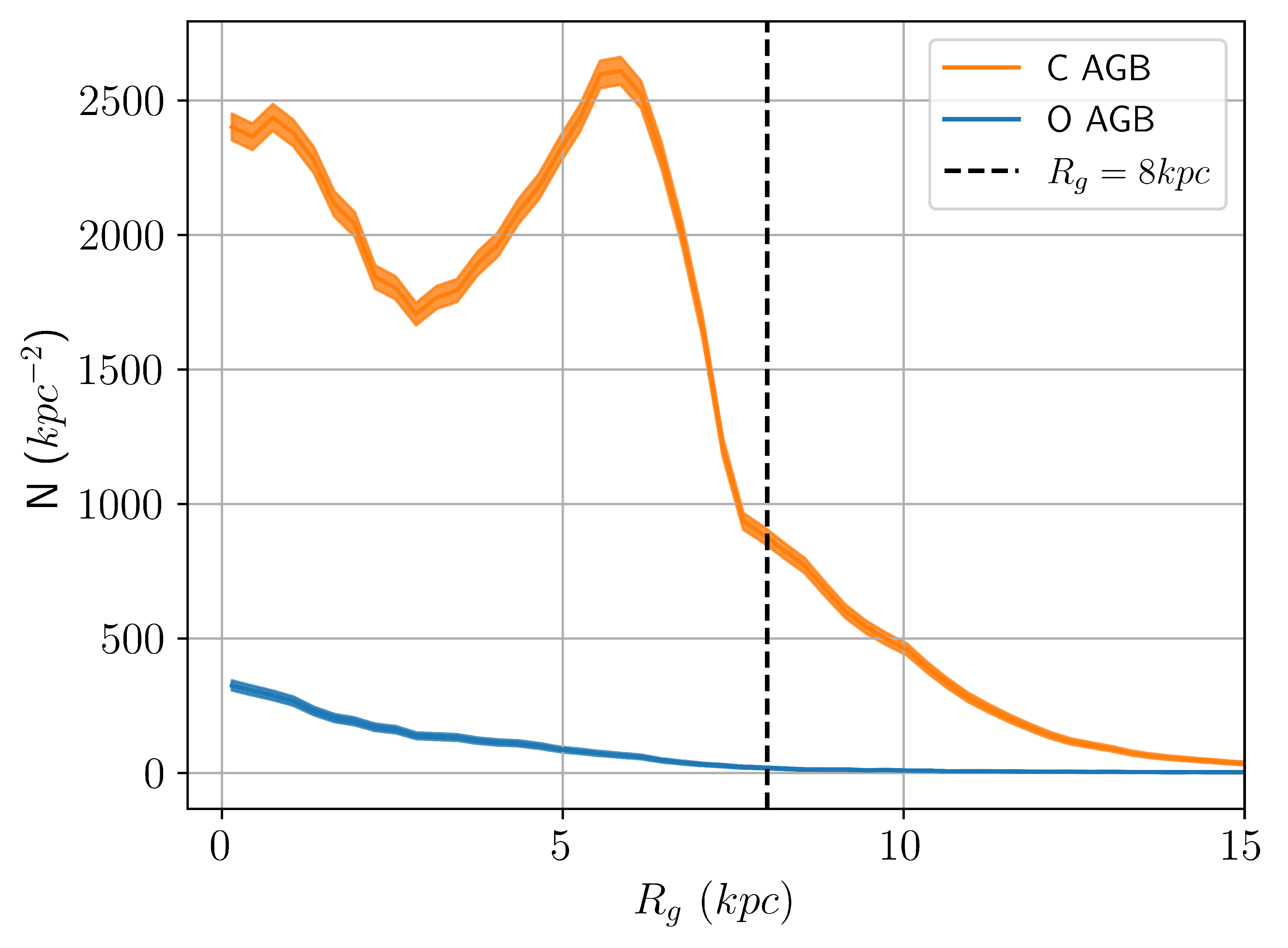}
    \caption{Plot of variation in areal density of C and O~AGB stars located close to the Galactic plane ($|b| < 10^\circ$) as a function of Galactocentric distance ($R_g$). The stripes indicate the
uncertainty in star count due to Poisson statistics. $R_g$ is estimated using \textit{Gaia}-DR3 parallax and law of cosines assuming the 8\,kpc as the distance to the Galactic center. The areal density at each $R_g$ is computed by counting the number of stars in an annulus of width 0.3\,kpc about $R_g$ and dividing the star count by the corresponding area.}
    \label{fig:N_vs_Rg_AGB_stars}
\end{figure}

%%%%%%%%%%%%%%%%%%%%%%%%%%%% FIG 12 Ends %%%%%%%%%%%%%%%%%%%%%%%%%%%%%%%%%%%%%%%%%%%

Similar to the analysis carried out for YSOs towards the Cygnus-X region, we considered the \textit{Gaia} counterparts to the identified C and O~AGB stars. Among the sources with \textit{Gaia} associations, we consider all sources whose parallax uncertainties are listed and are close to the Galactic plane ($|b| < 10^\circ$). We do not put any constraint on the uncertainty values of parallax as we are interested in the statistical distribution of these sources across the Galaxy, and including constraints in uncertainties drastically reduces the number of \textit{Gaia} associations. The former sample includes 550,930 C~AGB and 24,983 O~AGB stars. 

The areal densities of C and O~AGB stars close to the Galactic plane as a function of Galactocentric distance ($R_g$) are shown in Fig.~\ref{fig:N_vs_Rg_AGB_stars}, which indicates that for the identified C and O~AGB stars in the target list, the areal density of C~AGB stars dominate over that of O~AGB stars at all Galactocentric radii. Moreover, we find a peak in areal density of C~AGB stars at $R_g\simeq 6$\,kpc. We expect that these C~AGB stars in the peak are likely to be located at $\sim 2$\,kpc from the Sun, owing to the brightness criteria applied to the target list. A histogram of distances (derived from \textit{Gaia} parallax) to the C~AGB stars from the Sun with $5$\,kpc $<R_g<7$\,kpc indicates that the distribution peaks indeed at $\simeq 2.7$\,kpc as expected. This instils confidence in our classifications of C and O~AGB stars and implies that our classifier predictions of C and O~AGB stars are quite reliable. Thus, one can get a genuine list of candidate C and O~AGB stars in the target list by applying a threshold of 90\% on the output probability (in each stage of the classifier). 

\section{Summary}
\label{sec:summary}

The main motivation of this work is to classify the sources in the SPHEREx target list, and for this we have collated the training data for the machine learning algorithms from various published catalogs of sources of YSOs, AGN, MS and AGB stars. We used three supervised machine learning algorithms that give probability distribution over output classes, namely RF, FNN and CNN, to build an ensemble classifier. We have employed two-stage classification for YSOs and AGB stars. Thus, each source was probabilistically classified into one of the eight classes, including AGN, C and O~AGB stars, (reddened) MS stars, and YSO classes (Class I, II, flat~spectrum, and Class III) by considering the output probabilities from both the stages. We filter the targets with and without reliable photometry in $W3$ and $W4$ bands and classify those with real photometry in these bands using a 2-stage ensemble classifier which uses photometry in 2MASS $J$, $H$, $K_s$ and AllWISE $W1$, $W2$, $W3$ and $W4$ bands for classification. Whereas, those targets without good photometry in $W3$ and $W4$ bands are classified using another 2-stage ensemble classifier which uses only the photometry in 2MASS $J$, $H$, $K_s$ and AllWISE $W1$ and $W2$ bands for classification.

Our trained two-stage classifier was applied to the target list and we found that out of 8,308,384 target list sources, 1,966,340 were classified with a probability exceeding 90\% by the stage-1 classifier. Among them $1.7\%$ are YSOs, $58.2\%$ are AGB stars, $40.0\%$ are (reddened) MS stars whereas AGN constitute merely 2005 sources. The spatial distribution of these targets indicates that the YSOs are located close to the Galactic plane and in some higher latitude molecular clouds. Most of the identified AGB stars and {reddened} MS stars are located in the inner Galaxy.

We validate our classification by analysing the distributions of YSOs towards the star-forming complex Cygnus-X, where we find that the candidate YSOs distribution matches the filamentary structure of the cloud complex. The YSOs having \textit{Gaia} associations are found at distances $1.76\pm0.56$\,kpc, which is consistent with the distance to the complex. The areal density distribution of C and O~AGB stars across the Galaxy is analysed and we find consistent results. Thus, we can conclude that reliable class predictions can be obtained from our classifier with suitable threshold applied on the output probability.

Finally, although the class labels are predicted for sources in the target list, decisive classification requires further analysis through spectroscopy and/or alternate methods. 

\section*{Data Availability}

The table of results obtained after classifying the sources in the SPHEREx target list is included in SPLICES catalog and published online in machine readable format through IRSA.

\section*{Acknowledgements}

This research has made use of the NASA/IPAC Infrared Science Archive, which is funded by the National Aeronautics and Space Administration and operated by the California Institute of Technology. This publication makes use of data products from the Wide-Field Infrared Survey Explorer, which is a joint project of the University of California, Los Angeles, and the Jet Propulsion Laboratory/California Institute of Technology, funded by the National Aeronautics and Space Administration. This publication also makes use of data products from 2MASS, which is a joint project of the University of Massachusetts and the Infrared Processing and Analysis Center/California Institute of Technology, funded by the National Aeronautics and Space Administration and the National Science Foundation. 
This research has made use of the VizieR catalogue access tool, CDS, Strasbourg, France (DOI : 10.26093/cds/vizier). The original description of the VizieR service was published in 2000, A\&AS 143, 23. This work has made use of the SIMBAD database, operated at CDS, Strasbourg, France. This work has made use of data from the European Space Agency (ESA) mission
{\it Gaia} (\url{https://www.cosmos.esa.int/gaia}), processed by the {\it Gaia} Data Processing and Analysis Consortium (DPAC, \url{https://www.cosmos.esa.int/web/gaia/dpac/consortium}). Funding for the DPAC has been provided by national institutions, in particular the institutions participating in the {\it Gaia} Multilateral Agreement.

%%%%%%%%%%%%%%%%%%%%%%%%%%%%%%%%%%%%%%%%%%%%%%%%%%



\bibliographystyle{mnras}
\bibliography{YSO_classification_refs}

%%%%%%%%%%%%%%%%%%%%%%%%%%%%%%%%%%%%%%%%%%%%%%%%%%

%%%%%%%%%%%%%%%%% APPENDICES %%%%%%%%%%%%%%%%%%%%%

\appendix

\section{Assembled Training Data References (catalogues)}\label{sec:training_data_refs}

Mira Variables \citep{chen}, C and O~AGB \citep{suh}, AGN \citep{assef}, YSO \citep{{fischer}, {grobschedl}, {allen}, {megeath}, {kuhn}, {szegedi-elek}, {rebollido}, {pascucci}, {kun2009}, {kun2016}, {kumar}, {kim2016}, {frasca}, {fang}, {cooper}, {dzib}, {connelley}, {alcala}, {oliveira}, {ansdell2016}, {erickson}, {ansdell2017}}.

%%%%%%%%%%%%%%%%%%%%%%%%%%%%%%%%%%%%%%%%%%%%%%%%%%

\section{Ensemble helps reduce effects of over-fitting}
\label{sec:ensemble_over_fitting}

Over-fitting commonly occurs when there are insufficient samples for training and the model complexity is high. Popular ways to reduce over-fitting include using different regularization methods, reducing model complexity, increasing the number of training samples, adding noise to input features and using ensemble of classifiers. To reduce the effects of over-fitting we use an ensemble of different base classifiers. Also, to prevent the classifiers from overconfidently classifying the sources, we adopt the uncertainty based data resampling method. 

The ensemble approach is helpful in getting reliable output probabilities, which clearly represent the ambiguity faced by the classifier. Different class objects with similar features tend to be close to the separating boundary in the feature space. An over-fit classifier would tend to over-confidently predict the class label based on its distance from the learned decision boundary into one or the other class. On the other hand, an ensemble classifier reflects the inherent ambiguity in classification in the form of low output probabilities. 

The decision boundaries learned by each over-fit base classifier are different (since different subset of the training data is used for building each base classifier during cross-validation). Hence, each over-fit base classifier might predict different class to the same source. In that case, the ensemble classifier gives a relatively low probability of the source being associated with each output class. Contrarily, if a majority of the base classifiers predict the same class label for a source with relatively high probability, then the ensemble classifier also associates the source to that class with a relatively high probability. This way, the ensemble approach aids in obtaining reliable output probabilities, that can be used for applying suitable probability threshold and getting reliable class predictions.

\section{Cross-validation results}
\label{sec:cross-validation_results}

This section includes the cross-validation results for each type of base classifiers. The confusion matrix is determined for each type of classifier in each stage to gain insight into the confusion faced by the trained classifier. A confusion matrix denotes the number of known sources of a particular type, say `A', being classified as type `B' by the classifier, where `A' and `B' can be any of the output classes. Among the sources predicted as belonging to a particular class by the classifier, the fraction of them which are correctly classified is called precision (or purity). Out of known sources of a particular class, the fraction of them which are correctly classified by the classifier is called recall (or recovery rate). Both precision and recall are class-specific metrics. Here, the confusion matrices are normalized to display the recovery rate (recall) for each class as diagonal entries and the percentage of samples mis-classified into other output classes as off-diagonal entries. The purity of classification for each class is denoted as precision scores in the confusion matrices. For the $non-W3W4$ ensemble classifier stages-1 and 2, the confusion matrices are displayed in Tables~\ref{tab:4classes_ensemble} and \ref{tab:AGBclasses_ensemble}, and for the $W3W4$ ensemble classifier stages-1 and 2, the confusion matrices are displayed in Tables~\ref{tab:4classes_ensemble_W34} and \ref{tab:AGBclasses_ensemble_W34}. For RF in stages-1 and 2 of the $non-W3W4$ classifier, the confusion matrices are displayed in Tables~\ref{tab:4classes_rf} and \ref{tab:AGBclasses_rf} and that in stages-1 and 2 of the $W3W4$ classifier, the confusion matrices are displayed in Tables~\ref{tab:4classes_rf_W34} and \ref{tab:AGBclasses_rf_W34}. For FNN in stages-1 and 2 of the $non-W3W4$ classifier, the confusion matrices are displayed in Tables~\ref{tab:4classes_fcnn} and \ref{tab:AGBclasses_fcnn} and that in stages-1 and 2 of the $W3W4$ classifier, the confusion matrices are displayed in Tables~\ref{tab:4classes_fcnn_W34} and \ref{tab:AGBclasses_fcnn_W34}. For CNN in stages-1 and 2 of the $non-W3W4$ classifier, the confusion matrices are displayed in Tables~\ref{tab:4classes_cnn} and \ref{tab:AGBclasses_cnn} and that in stages-1 and 2 of the $W3W4$ classifier, the confusion matrices are displayed in Tables~\ref{tab:4classes_cnn_W34} and \ref{tab:AGBclasses_cnn_W34}. The number of samples of each class considered in the validation set for estimating the performance metrics is mentioned under the column of `Support'.

%%%%%%%%%%%%%%%%%%%%%%%%%%%% Table C1 Begins %%%%%%%%%%%%%%%%%%%%%%%%%%%%%%%%%%%%%%%%%%%

\begin{table}
    \centering
    	\caption{Normalized confusion matrix of the $non-W3W4$ ensemble classifier stage-1 4-class classifier with percentage values.}
    \label{tab:4classes_ensemble}
\begin{tabular}{l|l|c|c|c|c|c|}
\multicolumn{2}{c}{}&\multicolumn{4}{c}{Predicted Label} \\
\cline{2-7}
\multicolumn{1}{c|}{} & Class & AGB & AGN & MS & YSO & Support \\
\cline{2-7}
\multirow{5}{*}{\rotatebox[origin=c]{90}{True Label}} & AGB & 97.4 & 0.0 & 0.7 & 1.9 & $292980$ \\
\cline{2-7}
& AGN & 0.0 & 96.3 & 0.0 & 3.7 & $97220$ \\
\cline{2-7}
& MS & 0.1 & 0.0 & 96.1 & 3.8 & $341280$ \\
\cline{2-7}
& YSO & 2.2 & 2.4 & 7.4 & 88.0 & $341280$ \\
\cline{2-7}
& Precision & 97.4 & 91.9 & 92.2 & 93.1 & \multicolumn{1}{c}{}\\
\cline{2-6}
\end{tabular}
\end{table}

%%%%%%%%%%%%%%%%%%%%%%%%%%%% Table C1 Ends %%%%%%%%%%%%%%%%%%%%%%%%%%%%%%%%%%%%%%%%%%%

%%%%%%%%%%%%%%%%%%%%%%%%%%%% Table C2 Begins %%%%%%%%%%%%%%%%%%%%%%%%%%%%%%%%%%%%%%%%%%%

\begin{table}
    \centering
	\caption{Normalized confusion matrix of the $non-W3W4$ ensemble classifier stage-2 AGB sub-classifier with percentage values.}
    \label{tab:AGBclasses_ensemble}
\begin{tabular}{l|l|c|c|c|}
\multicolumn{2}{c}{}&\multicolumn{2}{c}{Predicted Label} \\
\cline{2-5}
\multirow{3}{*}{\rotatebox[origin=c]{90}{True Label}} & Class & C AGB & O AGB & Support \\
\cline{2-5}
\multicolumn{1}{c|}{} & C AGB & 78.8 & 21.2 & $44140$ \\
\cline{2-5}
& O AGB & 10.8 & 89.2 & $111340$ \\
\cline{2-5}
& Precision & 74.3 & 91.4 & \multicolumn{1}{c}{}\\
\cline{2-4}
\end{tabular}
\end{table}

%%%%%%%%%%%%%%%%%%%%%%%%%%%% Table C2 Ends %%%%%%%%%%%%%%%%%%%%%%%%%%%%%%%%%%%%%%%%%%%

%%%%%%%%%%%%%%%%%%%%%%%%%%%% Table C3 Begins %%%%%%%%%%%%%%%%%%%%%%%%%%%%%%%%%%%%%%%%%%%

\begin{table}
    \centering
    	\caption{Normalized confusion matrix of the $W3W4$ ensemble classifier stage-1 4-class classifier with percentage values.}
    \label{tab:4classes_ensemble_W34}
\begin{tabular}{l|l|c|c|c|c|c|}
\multicolumn{2}{c}{}&\multicolumn{4}{c}{Predicted Label} \\
\cline{2-7}
\multicolumn{1}{c|}{} & Class & AGB & AGN & MS & YSO & Support \\
\cline{2-7}
\multirow{5}{*}{\rotatebox[origin=c]{90}{True Label}} & AGB & 99.1 & 0.0 & 0.2 & 0.8 & $262960$ \\
\cline{2-7}
& AGN & 0.0 & 96.8 & 0.1 & 3.1 & $91220$ \\
\cline{2-7}
& MS & 0.4 & 0.1 & 82.6 & 16.9 & $25780$ \\
\cline{2-7}
& YSO & 0.9 & 2.5 & 7.5 & 89.0 & $122700$ \\
\cline{2-7}
& Precision & 99.5 & 96.6 & 68.7 & 92.2 & \multicolumn{1}{c}{}\\
\cline{2-6}
\end{tabular}
\end{table}

%%%%%%%%%%%%%%%%%%%%%%%%%%%% Table C3 Ends %%%%%%%%%%%%%%%%%%%%%%%%%%%%%%%%%%%%%%%%%%%

%%%%%%%%%%%%%%%%%%%%%%%%%%%% Table C4 Begins %%%%%%%%%%%%%%%%%%%%%%%%%%%%%%%%%%%%%%%%%%%

\begin{table}
    \centering
	\caption{Normalized confusion matrix of the $W3W4$ ensemble classifier stage-2 AGB sub-classifier with percentage values.}
    \label{tab:AGBclasses_ensemble_W34}
\begin{tabular}{l|l|c|c|c|}
\multicolumn{2}{c}{}&\multicolumn{2}{c}{Predicted Label} \\
\cline{2-5}
\multirow{3}{*}{\rotatebox[origin=c]{90}{True Label}} & Class & C AGB & O AGB & Support \\
\cline{2-5}
\multicolumn{1}{c|}{} & C AGB & 94.3 & 5.7 & $32120$ \\
\cline{2-5}
& O AGB & 2.2 & 97.8 & $103340$ \\
\cline{2-5}
& Precision & 92.9 & 98.2 & \multicolumn{1}{c}{}\\
\cline{2-4}
\end{tabular}
\end{table}

%%%%%%%%%%%%%%%%%%%%%%%%%%%% Table C4 Ends %%%%%%%%%%%%%%%%%%%%%%%%%%%%%%%%%%%%%%%%%%%

%%%%%%%%%%%%%%%%%%%%%%%%%%%% Table C5 Begins %%%%%%%%%%%%%%%%%%%%%%%%%%%%%%%%%%%%%%%%%%%

\begin{table}
    \centering
    	\caption{Normalized confusion matrix of the $non-W3W4$ RF in stage-1 4-class classifier with percentage values.}
    \label{tab:4classes_rf}
\begin{tabular}{l|l|c|c|c|c|c|}
\multicolumn{2}{c}{}&\multicolumn{4}{c}{Predicted Label} \\
\cline{2-7}
\multicolumn{1}{c|}{} & Class & AGB & AGN & MS & YSO & Support \\
\cline{2-7}
\multirow{5}{*}{\rotatebox[origin=c]{90}{True Label}} & AGB & 97.5 & 0.0 & 0.7 & 1.8 & 292980 \\
\cline{2-7}
& AGN & 0.0 & 94.5 & 0.0 & 5.5 & 97220 \\
\cline{2-7}
& MS & 0.1 & 0.0 & 95.8 & 4.1 & 341280 \\
\cline{2-7}
& YSO & 1.9 & 1.4 & 6.9 & 89.9 & 341280 \\
\cline{2-7}
& Precision & 97.7 & 95.2 & 92.8 & 92.6 & \multicolumn{1}{c}{}\\
\cline{2-6}
\end{tabular}
\end{table}

%%%%%%%%%%%%%%%%%%%%%%%%%%%% Table C5 Ends %%%%%%%%%%%%%%%%%%%%%%%%%%%%%%%%%%%%%%%%%%%

%%%%%%%%%%%%%%%%%%%%%%%%%%%% Table C6 Begins %%%%%%%%%%%%%%%%%%%%%%%%%%%%%%%%%%%%%%%%%%%

\begin{table}
    \centering
	\caption{Normalized confusion matrix of the $non-W3W4$ RF in stage-2 AGB sub-classifier with percentage values.}
    \label{tab:AGBclasses_rf}
\begin{tabular}{l|l|c|c|c|}
\multicolumn{2}{c}{}&\multicolumn{2}{c}{Predicted Label} \\
\cline{2-5}
\multirow{3}{*}{\rotatebox[origin=c]{90}{True Label}} & Class & C AGB & O AGB & Support \\
\cline{2-5}
\multicolumn{1}{c|}{} & C AGB & 72.2 & 27.8 & 44140 \\
\cline{2-5}
& O AGB & 6.5 & 93.5 & 111340 \\
\cline{2-5}
& Precision & 81.4 & 89.4 & \multicolumn{1}{c}{}\\
\cline{2-4}
\end{tabular}
\end{table}

%%%%%%%%%%%%%%%%%%%%%%%%%%%% Table C6 Ends %%%%%%%%%%%%%%%%%%%%%%%%%%%%%%%%%%%%%%%%%%%

%%%%%%%%%%%%%%%%%%%%%%%%%%%% Table C7 Begins %%%%%%%%%%%%%%%%%%%%%%%%%%%%%%%%%%%%%%%%%%%

\begin{table}
    \centering
    	\caption{Normalized confusion matrix of the $W3W4$ RF stage-1 in 4-class classifier with percentage values.}
    \label{tab:4classes_rf_W34}
\begin{tabular}{l|l|c|c|c|c|c|}
\multicolumn{2}{c}{}&\multicolumn{4}{c}{Predicted Label} \\
\cline{2-7}
\multicolumn{1}{c|}{} & Class & AGB & AGN & MS & YSO & Support \\
\cline{2-7}
\multirow{5}{*}{\rotatebox[origin=c]{90}{True Label}} & AGB & 99.4 & 0.0 & 0.0 & 0.5 & 262960 \\
\cline{2-7}
& AGN & 0.0 & 96.4 & 0.0 & 3.6 & 91220 \\
\cline{2-7}
& MS & 0.8 & 0.0 & 69.9 & 29.4 & 25780 \\
\cline{2-7}
& YSO & 0.9 & 1.7 & 1.8 & 95.5 & 122700 \\
\cline{2-7}
& Precision & 99.5 & 97.6 & 88.7 & 90.5 & \multicolumn{1}{c}{}\\
\cline{2-6}
\end{tabular}
\end{table}

%%%%%%%%%%%%%%%%%%%%%%%%%%%% Table C7 Ends %%%%%%%%%%%%%%%%%%%%%%%%%%%%%%%%%%%%%%%%%%%

%%%%%%%%%%%%%%%%%%%%%%%%%%%% Table C8 Begins %%%%%%%%%%%%%%%%%%%%%%%%%%%%%%%%%%%%%%%%%%%

\begin{table}
    \centering
	\caption{Normalized confusion matrix of the $W3W4$ RF in stage-2 AGB sub-classifier with percentage values.}
    \label{tab:AGBclasses_rf_W34}
\begin{tabular}{l|l|c|c|c|}
\multicolumn{2}{c}{}&\multicolumn{2}{c}{Predicted Label} \\
\cline{2-5}
\multirow{3}{*}{\rotatebox[origin=c]{90}{True Label}} & Class & C AGB & O AGB & Support \\
\cline{2-5}
\multicolumn{1}{c|}{} & C AGB & 91.7 & 8.3 & 32120 \\
\cline{2-5}
& O AGB & 1.3 & 98.7 & 103340 \\
\cline{2-5}
& Precision & 95.5 & 97.4 & \multicolumn{1}{c}{}\\
\cline{2-4}
\end{tabular}
\end{table}

%%%%%%%%%%%%%%%%%%%%%%%%%%%% Table C8 Ends %%%%%%%%%%%%%%%%%%%%%%%%%%%%%%%%%%%%%%%%%%%

%%%%%%%%%%%%%%%%%%%%%%%%%%%% Table C9 Begins %%%%%%%%%%%%%%%%%%%%%%%%%%%%%%%%%%%%%%%%%%%

\begin{table}
    \centering
    	\caption{Normalized confusion matrix of the $non-W3W4$ FNN in stage-1 4-class classifier with percentage values.}
    \label{tab:4classes_fcnn}
\begin{tabular}{l|l|c|c|c|c|c|}
\multicolumn{2}{c}{}&\multicolumn{4}{c}{Predicted Label} \\
\cline{2-7}
\multicolumn{1}{c|}{} & Class & AGB & AGN & MS & YSO & Support \\
\cline{2-7}
\multirow{5}{*}{\rotatebox[origin=c]{90}{True Label}} & AGB & 97.2 & 0.0 & 0.8 & 2.0 & 292980 \\
\cline{2-7}
& AGN & 0.0 & 96.9 & 0.0 & 3.1 & 97220 \\
\cline{2-7}
& MS & 0.1 & 0.0 & 96.0 & 3.9 & 341280 \\
\cline{2-7}
& YSO & 2.2 & 3.1 & 7.5 & 87.2 & 341280 \\
\cline{2-7}
& Precision & 97.3 & 89.9 & 92.2 & 93.1 & \multicolumn{1}{c}{}\\
\cline{2-6}
\end{tabular}
\end{table}

%%%%%%%%%%%%%%%%%%%%%%%%%%%% Table C9 Ends %%%%%%%%%%%%%%%%%%%%%%%%%%%%%%%%%%%%%%%%%%%

%%%%%%%%%%%%%%%%%%%%%%%%%%%% Table C10 Begins %%%%%%%%%%%%%%%%%%%%%%%%%%%%%%%%%%%%%%%%%%%

\begin{table}
    \centering
	\caption{Normalized confusion matrix of the $non-W3W4$ FNN in stage-2 AGB sub-classifier with percentage values.}
    \label{tab:AGBclasses_fcnn}
\begin{tabular}{l|l|c|c|c|}
\multicolumn{2}{c}{}&\multicolumn{2}{c}{Predicted Label} \\
\cline{2-5}
\multirow{3}{*}{\rotatebox[origin=c]{90}{True Label}}& Class & C AGB & O AGB & Support \\
\cline{2-5}
\multicolumn{1}{c|}{} & C AGB & 83.1 & 16.9 & 44140 \\
\cline{2-5}
& O AGB & 16.6 & 83.4 & 111340 \\
\cline{2-5}
& Precision & 66.5 & 92.6 & \multicolumn{1}{c}{}\\
\cline{2-4}
\end{tabular}
\end{table}

%%%%%%%%%%%%%%%%%%%%%%%%%%%% Table C10 Ends %%%%%%%%%%%%%%%%%%%%%%%%%%%%%%%%%%%%%%%%%%%

%%%%%%%%%%%%%%%%%%%%%%%%%%%% Table C11 Begins %%%%%%%%%%%%%%%%%%%%%%%%%%%%%%%%%%%%%%%%%%%

\begin{table}
    \centering
    	\caption{Normalized confusion matrix of the $W3W4$ FNN in stage-1 4-class classifier with percentage values.}
    \label{tab:4classes_fcnn_W34}
\begin{tabular}{l|l|c|c|c|c|c|}
\multicolumn{2}{c}{}&\multicolumn{4}{c}{Predicted Label} \\
\cline{2-7}
\multicolumn{1}{c|}{} & Class & AGB & AGN & MS & YSO & Support \\
\cline{2-7}
\multirow{5}{*}{\rotatebox[origin=c]{90}{True Label}} & AGB & 98.6 & 0.0 & 0.4 & 0.9 & 262960 \\
\cline{2-7}
& AGN & 0.0 & 96.2 & 0.2 & 3.5 & 91220 \\
\cline{2-7}
& MS & 0.2 & 0.0 & 87.1 & 12.6 & 25780 \\
\cline{2-7}
& YSO & 0.9 & 3.1 & 13.8 & 82.2 & 122700 \\
\cline{2-7}
& Precision & 99.5 & 95.8 & 55.2 & 91.9 & \multicolumn{1}{c}{}\\
\cline{2-6}
\end{tabular}
\end{table}

%%%%%%%%%%%%%%%%%%%%%%%%%%%% Table C11 Ends %%%%%%%%%%%%%%%%%%%%%%%%%%%%%%%%%%%%%%%%%%%

%%%%%%%%%%%%%%%%%%%%%%%%%%%% Table C12 Begins %%%%%%%%%%%%%%%%%%%%%%%%%%%%%%%%%%%%%%%%%%%

\begin{table}
    \centering
	\caption{Normalized confusion matrix of the $W3W4$ FNN in stage-2 AGB sub-classifier with percentage values.}
    \label{tab:AGBclasses_fcnn_W34}
\begin{tabular}{l|l|c|c|c|}
\multicolumn{2}{c}{}&\multicolumn{2}{c}{Predicted Label} \\
\cline{2-5}
\multirow{3}{*}{\rotatebox[origin=c]{90}{True Label}}& Class & C AGB & O AGB & Support \\
\cline{2-5}
\multicolumn{1}{c|}{} & C AGB & 95.0 & 5.0 & 32120 \\
\cline{2-5}
& O AGB & 3.2 & 96.8 & 103340 \\
\cline{2-5}
& Precision & 90.2 & 98.4 & \multicolumn{1}{c}{}\\
\cline{2-4}
\end{tabular}
\end{table}

%%%%%%%%%%%%%%%%%%%%%%%%%%%% Table C12 Ends %%%%%%%%%%%%%%%%%%%%%%%%%%%%%%%%%%%%%%%%%%%

%%%%%%%%%%%%%%%%%%%%%%%%%%%% Table C13 Begins %%%%%%%%%%%%%%%%%%%%%%%%%%%%%%%%%%%%%%%%%%%

\begin{table}
    \centering
    	\caption{Normalized confusion matrix of the $non-W3W4$ CNN in stage-1 4-class classifier with percentage values.}
    \label{tab:4classes_cnn}
\begin{tabular}{l|l|c|c|c|c|c|}
\multicolumn{2}{c}{}&\multicolumn{4}{c}{Predicted Label} \\
\cline{2-7}
\multicolumn{1}{c|}{} & Class & AGB & AGN & MS & YSO & Support \\
\cline{2-7}
\multirow{5}{*}{\rotatebox[origin=c]{90}{True Label}} & AGB & 96.5 & 0.0 & 0.8 & 2.7 & 292980 \\
\cline{2-7}
& AGN & 0.0 & 96.2 & 0.0 & 3.8 & 97220\\
\cline{2-7}
& MS & 0.3 & 0.0 & 95.4 & 4.3 & 341280 \\
\cline{2-7}
& YSO & 2.9 & 4.0 & 8.1 & 85.0 & 341280 \\
\cline{2-7}
& Precision & 96.3 & 87.3 & 91.5 & 91.7 & \multicolumn{1}{c}{}\\
\cline{2-6}
\end{tabular}
\end{table}

%%%%%%%%%%%%%%%%%%%%%%%%%%%% Table C13 Ends %%%%%%%%%%%%%%%%%%%%%%%%%%%%%%%%%%%%%%%%%%%

%%%%%%%%%%%%%%%%%%%%%%%%%%%% Table C14 Begins %%%%%%%%%%%%%%%%%%%%%%%%%%%%%%%%%%%%%%%%%%%

\begin{table}
    \centering
	\caption{Normalized confusion matrix of the $non-W3W4$ CNN in stage-2 AGB sub-classifier with percentage values.}
    \label{tab:AGBclasses_cnn}
\begin{tabular}{l|l|c|c|c|}
\multicolumn{2}{c}{}&\multicolumn{2}{c}{Predicted Label} \\
\cline{2-5}
\multirow{3}{*}{\rotatebox[origin=c]{90}{True Label}} & Class & C AGB & O AGB & Support \\
\cline{2-5}
\multicolumn{1}{c|}{} & C AGB & 76.7 & 23.3 & 44140 \\
\cline{2-5}
& O AGB & 15.4 & 84.6 & 111340 \\
\cline{2-5}
& Precision & 66.4 & 90.2 & \multicolumn{1}{c}{}\\
\cline{2-4}
\end{tabular}
\end{table}

%%%%%%%%%%%%%%%%%%%%%%%%%%%% Table C14 Ends %%%%%%%%%%%%%%%%%%%%%%%%%%%%%%%%%%%%%%%%%%%

%%%%%%%%%%%%%%%%%%%%%%%%%%%% Table C15 Begins %%%%%%%%%%%%%%%%%%%%%%%%%%%%%%%%%%%%%%%%%%%

\begin{table}
    \centering
    	\caption{Normalized confusion matrix of the $W3W4$ CNN in stage-1 4-class classifier with percentage values.}
    \label{tab:4classes_cnn_W34}
\begin{tabular}{l|l|c|c|c|c|c|}
\multicolumn{2}{c}{}&\multicolumn{4}{c}{Predicted Label} \\
\cline{2-7}
\multicolumn{1}{c|}{} & Class & AGB & AGN & MS & YSO & Support \\
\cline{2-7}
\multirow{5}{*}{\rotatebox[origin=c]{90}{True Label}} & AGB & 93.3 & 0.1 & 1.7 & 4.9 & 262960 \\
\cline{2-7}
& AGN & 0.1 & 95.1 & 0.6 & 4.2 & 91220 \\
\cline{2-7}
& MS & 2.5 & 1.0 & 77.5 & 19.0 & 25780 \\
\cline{2-7}
& YSO & 3.8 & 5.4 & 16.6 & 74.1 & 122700 \\
\cline{2-7}
& Precision & 97.9 & 92.4 & 43.9 & 80.8 & \multicolumn{1}{c}{}\\
\cline{2-6}
\end{tabular}
\end{table}

%%%%%%%%%%%%%%%%%%%%%%%%%%%% Table C15 Ends %%%%%%%%%%%%%%%%%%%%%%%%%%%%%%%%%%%%%%%%%%%

%%%%%%%%%%%%%%%%%%%%%%%%%%%% Table C16 Begins %%%%%%%%%%%%%%%%%%%%%%%%%%%%%%%%%%%%%%%%%%%

\begin{table}
    \centering
	\caption{Normalized confusion matrix of the $W3W4$ CNN in stage-2 AGB sub-classifier with percentage values.}
    \label{tab:AGBclasses_cnn_W34}
\begin{tabular}{l|l|c|c|c|}
\multicolumn{2}{c}{}&\multicolumn{2}{c}{Predicted Label} \\
\cline{2-5}
\multirow{3}{*}{\rotatebox[origin=c]{90}{True Label}} & Class & C AGB & O AGB & Support \\
\cline{2-5}
\multicolumn{1}{c|}{} & C AGB & 93.1 & 6.9 & 32120 \\
\cline{2-5}
& O AGB & 3.9 & 96.1 & 103340 \\
\cline{2-5}
& Precision & 88.2 & 97.8 & \multicolumn{1}{c}{}\\
\cline{2-4}
\end{tabular}
\end{table}

%%%%%%%%%%%%%%%%%%%%%%%%%%%% Table C16 Ends %%%%%%%%%%%%%%%%%%%%%%%%%%%%%%%%%%%%%%%%%%%

% Don't change these lines
\bsp	% typesetting comment

\label{lastpage}
\end{document}